\documentclass[preprint2]{aastex701}

\newcommand{\mstar}{M_*}
\newcommand{\msun}{M_{\odot}}

\newcommand{\rsun}{R_\odot}
\newcommand{\mdot}{\dot{M}}
\newcommand{\numstars}{74}

\newcommand{\numslow}{55}
\newcommand{\teff}{T_{\text{eff}}}
\newcommand{\logg}{\log g}
\newcommand{\vinfty}{v_\infty}
\newcommand{\rnaught}{R_0}
\newcommand{\moperyear}{M_{\odot}\text{ yr}^{-1}}
\newcommand{\kms}{\text{km s}^{-1}}
\newcommand{\lstar}{L_*}

\newcommand{\halpha}{H$\alpha$}
\newcommand{\kpc}{\text{kpc}}
\newcommand{\arcsecs}{\text{arcsec}}
\newcommand{\Jysr}{\text{Jy sr}^{-1}}
\newcommand{\Jycmsrnuc}{\text{Jy cm}^2\text{ sr}^{-1}\text{nucleon}^{-1}}
\newcommand{\um}{\mu\text{m}}
\newcommand{\ergscmc}{\text{erg s}^{-1}\text{ cm}^{-2}/c}

\newcommand{\MJypersr}{\text{MJy sr}^{-1}}
\newcommand{\Jypersquarepix}{\text{Jy pix}^{-2}}

\usepackage{amsmath}

\begin{document}

\title{Mass-loss Rates of \numstars~Bow Shock OB Stars: No Evidence for Bi-stability Behavior}

\author[orcid=0009-0005-0582-8469,gname=Nikhil,sname=Patten]{Nikhil Patten}
\affiliation{Department of Physics and Astronomy, University of Wyoming, 1000 E. University Avenue, Laramie, WY 82071, USA}
\email[show]{npatten@uwyo.edu}

\author[orcid=0000-0002-4475-4176,gname=Henry,sname=Kobulnicky]{Henry A. Kobulnicky}
\affiliation{Department of Physics and Astronomy, University of Wyoming, 1000 E. University Avenue, Laramie, WY 82071, USA}
\email[show]{chipk@uwyo.edu}

\begin{abstract}

Accurate measurements of wind-driven mass-loss rates are essential for understanding the evolution and fate of massive stars. Established techniques for measuring $\dot{M}$ suffer from large dispersions and become insensitive in the regime of weak-winded late-O to early B stars. We employ a promising new technique to measure stellar mass loss for \numstars~early type stars that produce infrared  stellar bow shock nebulae. We use updated stellar parameters, improved parallactic distances and proper motions, and modern dust models to characterize the bow shock nebulae and obtain new measurements of mass-loss rates. $\mdot$ ranges from $2\times10^{-9}~\moperyear$ for weak-winded late-O and early B stars at $\log\left(\lstar/L_\odot\right)\approx4.5$ to $4\times10^{-6}~\moperyear$ for the most luminous stars in our sample at $\log\left(\lstar/L_\odot\right)\approx6.3$. Stellar space velocities relative to their local standard of rest are insufficient to accurately reflect the relative star-ISM velocities for the majority of our sample, which have speeds much less than canonical runaway thresholds, suggesting the pervasive influence of bulk interstellar flows in forming bow shock nebulae. Adopting a minimum relative velocity of $20~\kms$ is physically motivated and yields mass-loss rates in excellent agreement with the latest theoretical prescriptions while requiring no adjustable parameters. We find no enhancements in $\mdot$ for stars cooler than 25~kK, in agreement with modern mass-loss recipes that do not include bi-stability effects.


\end{abstract}


\section{Introduction}

Massive stars---stars with masses ranging from $8\lesssim\mstar\lesssim120~\msun$---are the sites of nucleosynthesis and the main sources of mechanical and radiative feedback in their host galaxies. Although the stellar initial mass function dictates that early type stars be far fewer in number than their low-mass counterparts, massive stars dominate the radiative output of their host galaxies. Due to their high luminosities, massive stars also host radiation-driven stellar winds. Although massive stars have other channels through which their mass is altered throughout their lifetimes (binary interactions being foremost among them), mass loss through stellar winds acts continually to reduce the star's mass and, when integrated over the star's lifetime, can reduce the star's mass by as much as one fourth \citep{CAK1975}. After the main- and post main-sequence stages of evolution, massive stars evolve into either neutron stars or black holes depending on the stellar mass at the moment of core collapse. Systems of degenerate objects are of particular importance in the age of gravitational wave astronomy \citep{Fryer2002, Pretorius2005, Kalogera2004,Abbott2016, Abbott2017a, Abbott2017b}.

Several techniques for measuring mass-loss rates have been developed since the discovery of massive star stellar outflows \citep{Puls2008, Smith2014, Vink2022}. Recent works combine multiple methods to obtain measurements of $\mdot$ while taking into account complicating factors \citep{Martins2005, Vink2023}. One common method makes use of the P-Cygni profiles in the UV metal-resonance transitions \citep{Garmany1981, HowarthandPrinja1989, Marcolino2009}. Two wind properties can be extracted from UV techniques---terminal wind speed $\vinfty$ and mass-loss rate $\mdot$. Deriving mass-loss rates $\mdot$ from UV resonance line technique requires modeling of parameters that are not easily observable, such as the clumping factor in the winds. Another challenge with UV techniques is that under typical conditions, UV resonance lines are saturated in absorption. In such cases, the derived mass-loss rate can only serve as a lower limit. Since the optical depth of the line is proportional to the density of the wind $\rho$, the UV method is classified as a $\rho$ diagnostic---a method in which the derived mass-loss rate scales linearly with the density of the wind.

Another technique to measure mass-loss rate utilizes the \halpha~recombination lines in the wind to measure mass-loss rates \citep{Leitherer1988, Lamers1993, Puls1996, Markova2004}. The benefit of \halpha~methods is that unlike the UV resonance line technique, there is no uncertainty about the ionization of hydrogen for hot stars, unlike metal species which exhibit several ionization states in different fractions. Since the \halpha-derived mass-loss rates scale as $\rho^2$, and hot stars are known to exhibit small-scale density inhomogeneities in their winds \citep[known as micro-clumping,][]{Feldmeir1997, RunacresandOwocki2002, DessartamdOwocki2005, Owocki2015}, it is common for \halpha~methods to yield large overestimates in $\mdot$. In their study of 40 Galactic O stars, \citet{Fullerton2006} compared $\rho$- and $\rho^2$-derived measurements of $\mdot$ and found a large discrepancy between the two methods. This discrepancy seemed to suggest microclumping was more significant than previously thought, and \halpha~methods would have to be revised significantly to account for these effects. However, once thought to be independent of clumping effects, UV methods may also suffer complications. Clumps that are optically thin in  \halpha\ may be optically thick at some frequencies in the UV \citep{Oskinova2007}---an effect known as ``macroclumping.'' Accounting for macroclumping in the winds can help resolve the discrepancy between H$\alpha$ and UV measuremens \citep{Surlan2013}. As of yet, the effects of wind micro- and macroclumping on mass loss and spectral transitions are still being investigated and a fully consistent study is still needed \citep{Vink2022}.








A new technique to measure mass-loss rates that is physically independent of existing methods uses stellar bow shock nebulae \citep[][hereafter K18 and K19, respectively]{K18, K19}. Stellar bow shock nebulae are arc-shaped features formed when dust grains in the interstellar medium are shocked and heated by the winds of a high-velocity, early type star. The mass-loss rate of the central star can be measured given information on the properties of the ambient interstellar medium and morphology of the resulting bow shock. \citet{GullandSofia1979} were the first to propose this technique in their work on the newly discovered shock-like nebulae around a T Orionis K0e variable star, LL Ori, and an O9.5 V runaway, $\zeta$ Oph, from optical images. Balancing ram pressure from the incident stellar wind on the bulk motion of the surrounding material, they found that the mass-loss rate of the central star could be calculated from

\begin{equation}
    \dot{M} = \frac{4\pi R_0^2v_*^2\rho_a}{v_w}\text{ ,}
    \label{eq:one}
\end{equation}

\noindent with $R_0$ representing the ``standoff distance'' between the star and the apex of the bow shock, $v_*$ being the velocity of the star with respect to its surrounding ISM, $\rho_a$ signifying the ambient density of the pre-shock ISM, and $v_w$ denoting the speed of the outflowing stellar wind. Using estimates on these parameters for both bow shock nebulae, \citet{GullandSofia1979} derived independent measures of mass-loss rates for both stars. In the case of $\zeta$ Oph, their value of $2.3\times10^{-8}~\moperyear$ agrees well with the contemporary \citep[][$2.20\times10^{-8}~\moperyear$]{Abbott1978} and modern \citep[][$1.91\times10^{-8}~\moperyear$]{Bjorklund2023} theoretical expectations. Using bow shocks to measure mass-loss rates would remain an untapped concept for several decades until a reformulation from K18 allowed this technique to be extended to the growing collection of IR wavelength bow shock nebulae \citep{vanBuren1995, NoriegaCrespo1997,Povich2008, Gvaramadze2010,K10, Peri2012, Peri2015, K16}. K18 refined measurements of mass-loss rates by inferring interstellar number densities from archival infrared Herschel Space Observatory images \citep[most targets selected from Hi-GAL survey][]{Molinari2010}. K18 applied this method to measure the mass-loss rates for 20 OB stars that supported infrared bow shock nebulae. With this preliminary sample, they found that bow shock-derived mass-loss rates were situated between \halpha\ measurements (uncorrected for clumping) and UV measurements. After a promising demonstration of the bow shock method, K19 expanded the bow shock sample to 70 OB stars. They found mass-loss rates ranging from $\sim10^{-7}$ to $10^{-9}~\moperyear$ with good agreement between results from the bow shock method and UV measurements from \citet{Martins2005} for early type stars ($\teff>35$~kK), but showed disagreement for late-O stars with bow shock measurements 1.0--1.5 orders of magnitude greater than UV measurements. They also found that \halpha~measurements were greater than bowhshock measurements for late-O stars, but noted that many of the \halpha~measurements were upper limits on account of the insensitivity of the \halpha~method in this regime of weak-winded stars. Interestingly, they found that stars on the cool side of the bi-stability jump (i.e., with effective temperatures $T_{\text{eff}}\lesssim 25$~kK) showed evidence for the predicted enhancement, with $\mdot$\ orders of magnitude greater than the \citet{Vink2001} prescription, which included bi-stability enhancements. It is worth noting that their analysis was based on red wavelength optical spectra, which limited the precision on determinations of stellar effective temperature. For stars hotter than 25~kK, their results were $\sim0.4$~dex below the \citet{Vink2001} predictions, showing that, on average, bow shock-derived mass-loss rates are a factor of $3$ lower than the commonly used prescription.

\citet[][hereafter K22]{K22} studied the kinematics of stars at the center of bow shock nebulae. They found that although bow shock stars had a slightly higher median 2D space velocity than the control sample of O stars selected from the Galactic O-Star Catalog \citep[GOSC,][14.6~$\kms$ versus 11.3~$\kms$, respectively]{Maiz2013, Sota2014}, bow shock stars had an almost identical runaway fraction as the control sample (24\% versus 22\%). Said another way, bow shock stars are not exclusively runaway stars. In fact, the majority of bow shock stars have space velocities less than 25~$\kms$ (76\%). It is also true that not all runaway stars are bow shock stars. In their search of bow shock nebulae around runaway stars, \citet{Peri2015} found that only 10--15\% of runaway stars exhibit IR bow shocks. K18 proposed the interstellar density was the determining factor in forming an IR bow shock rather than the central star's spectral type or velocity with respect to its surroundings. They found an empirical lower limit on the ambient interstellar density of $n_a \gtrsim5~\text{cm}^{-3}$. Following this result, K22 identified two distinct populations of bow shock nebulae: fast-moving stars in which the morphological symmetry axis of the bow shock is parallel to the star's velocity vector (31\%) and slow-moving stars with an isotropic distribution of alignments between the bow shock's morphology and peculiar motion (69\%). They also noted that stars in the slow-moving, randomly aligned group tended to face \ion{H}{2} regions or bright-rimmed clouds visible at 8~$\um$---the ``in situ'' bow shocks, coined by \cite{Povich2008}. This led the authors to conclude that local flows played a role in forming the bow shock in these cases, while the aligned group arose from a scenario in which the star's runaway velocity generated the supersonic motion needed to form a bow shock.

The opacity of the star's atmosphere due to metal-resonance transitions is thought to play a significant role in the rate at which mass is expelled from the star through these stellar winds. Therefore, some works have predicated that recombination from one ion to another can produce dramatic increases in mass-loss rate $\mdot$---the bi-stability jump. Simulations of \citet{Vink2001} predict a bi-stability jump at stellar effective temperature $T_{\text{eff}}\simeq25$~kK due to the recombination of \ion{Fe}{4} to \ion{Fe}{3}, and minor bi-stability jumps at 15 and 35~kK, which the authors attributed to the recombination of \ion{Fe}{3} to \ion{Fe}{2} and \ion{C}{4} to \ion{C}{3}, respectively. Observational works have thus set out to find evidence of this predicted bi-stability jumps, focusing primarily on the bi-stability jump at $T_{\text{eff}}\simeq25$~kK. \citet{Benaglia2007}, for example, used radio-continuum techniques to study $\mdot$ for 12 late-O to early B supergiants and found some evidence of enhancements due to bi-stability. The previous bow shock work of K19 also reported increases $\mdot$ for stars cooler than 25~kK. However, recent works find no increase in mass-loss rates for stars on the cool side of the bi-stability gap \citep{deBurgos2024, Alkousa2025}.


In this work, we measure the mass-loss rates for \numstars~stars that lie at the center of bow shock nebulae in the same manner as K18/K19, with some important improvements. The Gaia Data Release 3 \citep[Gaia DR3,][]{GaiaDR32023} provides newer parallaxes and proper motions for bow shock-driving stars, yielding more accurate measurements of distances and peculiar space velocities. In addition, \citet[][P25]{Patten2025} used blue-optical spectra to obtain more precise measurements of stellar temperature $\teff$, surface gravity $\logg$, and projected-rotational velocity $v\sin i$ for a large sample of bow shock-driving stars. This improvement enables more accurate estimates of terminal wind speeds $\vinfty$, another factor in mass-loss-rate calculations. Finally, \citet{Whisnant2025} used citizen science data to tabulate the stand-off distances $\rnaught$ for a subset of stars in the \citet[][K16]{K16} and \citet[][J19]{J19} bow shock catalogs, from which this work selects targets. These combined improvements enable more robust estimates for a comparable number of early-type stars as K19.

In Section~\ref{sec:targs} we describe our target selection. In Section~\ref{sec:measuringmdot} we present the parameters needed to measure $\mdot$ using the bow shock method and how we obtain each parameter. In Section~\ref{sec:results} we present updated measurements of $\mdot$ and compare our values to different theoretical prescriptions. In Section~\ref{sec:discussion} we summarize the main findings of this work and discuss the implications of our results on the predicted bi-stability effect and what our results suggest about stars that form bow shocks and their velocities relative to their surroundings/






\section{Target Selection}
\label{sec:targs}

We selected stars with previously measured stellar parameters from the spectroscopic study P25, which selected a subset of bright, northern targets with reliable parallax measurements from the K16/J19 catalogs. We then limited these 103 P25 OB stars to only include stars with 70~$\um$ images from \textit{Herschel Space Observatory} (\textit{HSO}) surveys. This cut yielded 71 bow shock stars. We also searched the literature for K16/J19 stars with \textit{HSO} data that also have existing spectroscopic references in the literature. This led us to search the large-sclae spectroscopic O- and B-star project IACOB \citep{IACOBX} and a study on the Carina nebula \citep{HanesandPovich2018}. We then fitted the SED of these stars using the literature measurements on fundamental stellar parameters $\teff$ and $\logg$. This yielded three additional stars\footnote{The spectroscopic references for these three stars are as follows: BS$407_\star$ (HD 93249)---\citet{Holgado2018}; BS$409_\star$ (HD 93027)---\citet{Holgado2022}; BS$414_\star$ (V* V798 Car)---\citet{HanesandPovich2018}}, raising the total number to \numstars.



\begin{longrotatetable}
\begin{deluxetable*}{lcccccccccccccc}
	\tablecaption{Spectral and stellar parameters of program stars}\label{tab:BS_observed}
	\tabletypesize{\scriptsize}
	\tablehead{
		K16/J19 ID & Common Identifier & R.A. (2000) & Decl. (2000)& $T$ & $\sigma_T$ & $\log g$ & $\sigma_{\log g}$ & $R_{\star}$ & $\sigma_{R_\star}$ & $\log \left(L_\star/L_\odot\right)$ & $\sigma_{\log\left(\lstar/L_\odot \right)}$ & $M_\star$ & $\sigma_{M_\star}$ & binary \\ & & (HH:MM:SS) & ($^\circ$:$'$:$"$) & (kK) & (kK) & & &($R_\odot$) & ($R_\odot$) & (dex) & (dex) & ($M_\odot$) & ($M_\odot$) & \\
		(1) & (2) & (3) & (4) & (5) & (6) & (7) & (8) & (9) & (10) & (11) & (12) & (13) & (14) & (15)}
	\startdata
		 BS001$_\star$ & CD-29 14004 & 17:48:07.68 & -29:07:55.6 & 34.1 & 2.0 & 3.83 & 0.25 & 10.9 & 0.78 & 5.13 & 0.08 & 53.1 & 26.0 & SB2\\
		 BS007$_\star$ & ... & 17:58:30.65 & -26:09:49.3 & 22.3 & 1.5 & 4.44 & 0.18 & 3.57 & 0.2 & 3.44 & 0.10 & 7.8 & 0.84 & ...\\
		 BS010$_\star$ & HD 314937 & 18:00:17.47 & -25:14:14.6 & 27.8 & 1.5 & 3.81 & 0.16 & 8.03 & 0.51 & 4.44 & 0.08 & 13.6 & 1.2 & ...\\
		 BS013$_\star$ & $\zeta$ Oph & 16:37:09.55 & -10:34:01.2 & 32.1 & 0.5 & 3.7 & 0.06 & 8.46 & 0.47 & 4.82 & 0.06 & 14.9 & 0.9 & ...\\
		 BS016$_\star$ & ... & 18:00:55.22  & -22:57:37.1 & 27.4 & 1.9 & 3.36 & 0.18 & 16.8 & 3.2 & 5.19 & 0.25 & 34.8 & 9.1 & ...\\
		 BS017$_\star$ & ... & 18:03:33.49 & -22:34:38.3 & 25.3 & 2.1 & 4.04 & 0.3 & 22.5 & 8.3 & 5.34 & 0.24 & 87.0 & 27.0 & SB2\\
		 BS019$_\star$ & ... & 18:05:11.18 & -21:04:43.3 & 24.5 & 2.0 & 3.48 & 0.25 & 31.0 & 13.0 & 5.65 & 0.30 & 78.0 & 24.0 & ...\\
		 BS026$_\star$ & ... & 18:11:59.40 & -19:36:55.4 & 28.5 & 1.0 & 3.9 & 0.16 & 15.29 & 0.99 & 5.16 & 0.07 & 72.0 & 24.0 & SB2\\
		 BS027$_\star$ & TYC 6272-1457-1 & 18:12:13.03 & -19:35:24.0 & 22.5 & 1.0 & 3.96 & 0.13 & 5.46 & 1.2 & 3.86 & 0.26 & 9.4 & 1.8 & ...\\
		 BS043$_\star$ & TYC 6269-2270-1 & 18:16:25.25 & -17:05:20.8 & 26.9 & 1.1 & 3.74 & 0.13 & 8.25 & 0.77 & 4.53 & 0.10 & 14.2 & 1.1 & R\\
		 BS051$_\star$ & NGC 6618 258 & 18:20:22.70 & -16:08:34.1 & 34.0 & 1.0 & 4.41 & 0.21 & 10.9 & 1.7 & 5.10 & 0.13 & 95.0 & 30.0 & SB2\\
		 BS056$_\star$ & ... & 18:17:17.18 & -15:29:25.4 & 26.8 & 2.0 & 3.71 & 0.24 & 28.8 & 6.0 & 5.64 & 0.23 & 146.0 & 99.0 & SB2; R\\
		 BS061$_\star$ & HD 168183 & 18:18:58.70 & -13:59:28.3 & 30.6 & 0.6 & 3.83 & 0.12 & 15.3 & 1.4 & 5.26 & 0.08 & 64.9 & 17.0 & EB; R\\
		 BS063$_\star$ & BD-13 4934 & 18:19:05.57 & -13:54:50.4 & 31.1 & 1.1 & 4.21 & 0.21 & 7.08 & 0.42 & 4.59 & 0.06 & 58.0 & 39.0 & SB2\\
		 BS065$_\star$ & BD-13 4937 & 18:19:20.04 & -13:54:21.6 & 21.2 & 0.7 & 3.72 & 0.09 & 6.22 & 0.31 & 3.82 & 0.06 & 9.1 & 0.7 & ...\\
		 BS067$_\star$ & NGC 6611 584 & 18:18:23.64 & -13:36:28.1 & 32.2 & 1.2 & 4.15 & 0.21 & 8.54 & 0.5 & 4.85 & 0.07 & 59.6 & 15.0 & ...\\
		 BS150$_\star$ & TYC 5118-279-1 & 18:49:25.06 & -02:21:09.7 & 25.1 & 1.1 & 3.97 & 0.13 & 8.42 & 0.42 & 4.44 & 0.07 & 16.1 & 3.0 & R\\
		 BS200$_\star$ & ... & 19:02:08.86 & 3:30:47.2 & 25.1 & 1.8 & 3.5 & 0.25 & 32.8 & 9.4 & 5.57 & 0.31 & 80.0 & 23.0 & ...\\
		 BS255$_\star$ & TYC 1054-952-1 & 19:19:00.82 & 13:42:59.0 & 20.0 & 0.8 & 3.71 & 0.1 & 6.17 & 0.29 & 3.72 & 0.07 & 8.5 & 0.64 & ...\\
		 BS288$_\star$ & HD 350123 & 19:32:44.33 & 19:58:40.8 & 27.2 & 1.3 & 3.88 & 0.13 & 8.79 & 0.5 & 4.63 & 0.08 & 18.5 & 3.7 & ...\\
		 BS300$_\star$ & HD 344765 & 19:44:20.06 & 23:52:45.1 & 29.2 & 0.8 & 4.06 & 0.12 & 6.64 & 0.28 & 4.47 & 0.05 & 17.0 & 2.7 & ...\\
		 BS302$_\star$ & LS II +23 47 & 19:45:47.52 & 24:06:00.0 & 22.0 & 1.0 & 3.79 & 0.14 & 8.54 & 0.67 & 4.22 & 0.09 & 12.9 & 1.8 & R\\
		 BS303$_\star$ & ... & 19:46:06.17 & 24:11:13.9 & 20.2 & 1.6 & 2.7 & 0.18 & 68.4 & 22.0 & 5.90 & 0.31 & 85.0 & 21.0 & SB1\\
		 BS305$_\star$ & HD338936 & 19:46:22.68 & 24:37:48.0 & 33.5 & 1.0 & 4.59 & 0.15 & 10.2 & 1.5 & 5.08 & 0.15 & 93.0 & 25.0 & SB1; SB2; R\\
		 BS306$_\star$ & BD+24 3883 & 19:47:21.38 & 24:33:43.9 & 27.2 & 3.0 & 3.58 & 0.23 & 9.61 & 0.7 & 4.72 & 0.14 & 19.1 & 4.1 & SB2\\
		 BS314$_\star$ & HD 338961 & 19:50:24.41 & 27:27:55.8 & 26.5 & 1.4 & 3.69 & 0.15 & 12.4 & 1.5 & 4.85 & 0.15 & 31.0 & 33.0 & SB2\\
		 BS320$_\star$ & HD 191611 & 20:09:26.06 & 36:29:19.7 & 27.8 & 1.1 & 3.09 & 0.13 & 23.9 & 2.4 & 5.48 & 0.11 & 54.5 & 20.0 & ...\\
		 BS321$_\star$ & ... & 20:14:33.29 & 36:29:49.9 & 30.9 & 1.8 & 4.24 & 0.24 & 9.36 & 0.74 & 4.90 & 0.09 & 65.1 & 31.0 & ...\\
		 BS322$_\star$ & ... & 20:14:51.10 & 36:35:58.2 & 32.7 & 1.0 & 3.98 & 0.17 & 13.96 & 0.95 & 5.31 & 0.07 & 74.0 & 30.0 & SB1\\
		 BS324$_\star$ & LS II +38 19 & 20:13:28.92 & 38:14:31.6 & 32.4 & 1.0 & 4.01 & 0.15 & 17.3 & 1.3 & 5.49 & 0.08 & 114.0 & 49.0 & SB2\\
		 BS325$_\star$ & HD 194303 & 20:23:35.71 & 36:55:44.0 & 21.6 & 0.8 & 3.29 & 0.1 & 15.35 & 0.82 & 4.67 & 0.07 & 18.7 & 2.8 & R\\
		 BS330$_\star$ & HD 229159 & 20:22:54.02 & 39:12:28.8 & 22.4 & 1.0 & 2.94 & 0.12 & 28.4 & 1.4 & 5.25 & 0.07 & 30.9 & 5.5 & ...\\
		 BS331$_\star$ & LS II +39 53 & 20:27:17.57 & 39:44:32.6 & 34.7 & 1.0 & 3.92 & 0.13 & 16.9 & 0.43 & 4.84 & 0.05 & 89.0 & 1.1 & SB1; SB2\\
		 BS332$_\star$ & ... & 20:26:24.89 & 40:01:41.2 & 32.2 & 0.8 & 3.95 & 0.13 & 8.64 & 0.43 & 4.84 & 0.05 & 15.8 & 1.1 & SB1; SB2\\
		 BS335$_\star$ & TYC 3156-1106-1 & 20:28:15.38 & 40:44:04.6 & 24.4 & 1.2 & 3.32 & 0.14 & 16.99 & 0.77 & 4.92 & 0.05 & 21.0 & 2.2 & ...\\
		 BS339$_\star$ & ... & 20:36:04.51 & 40:56:13.2 & 45.6 & 4.5 & 4.28 & 0.33 & 16.1 & 5.0 & 6.00 & 0.36 & 89.0 & 24.0 & SB2\\
		 BS340$_\star$ & HD 195229 & 20:28:30.24 & 42:00:35.3 & 25.8 & 1.2 & 3.42 & 0.12 & 11.24 & 0.56 & 4.68 & 0.07 & 17.4 & 2.1 & ...\\
		 BS341$_\star$ & ... & 20:34:28.94 & 41:56:16.8 & 33.1 & 0.8 & 4.32 & 0.16 & 13.4 & 2.8 & 5.29 & 0.21 & 91.0 & 26.0 & SB1; SB2\\
		 BS344$_\star$ & BD+43 3654 & 20:33:36.07 & 43:59:07.4 & 46.8 & 0.9 & 3.784 & 0.05 & 21.69 & 0.87 & 6.31 & 0.04 & 104.0 & 12.0 & ...\\
		 BS350$_\star$ & ... & 21:22:55.70 & 50:24:25.2 & 24.2 & 1.6 & 3.16 & 0.17 & 16.5 & 4.5 & 4.90 & 0.23 & 22.1 & 6.3 & SB1; R\\
		 BS353$_\star$ & ... & 22:16:32.93 & 59:23:30.8 & 37.7 & 1.3 & 3.98 & 0.12 & 14.79 & 0.99 & 5.61 & 0.07 & 73.0 & 21.0 & SB2\\
		 BS356$_\star$ & HD 240015 & 22:39:17.71 & 59:01:00.5 & 26.2 & 1.0 & 3.25 & 0.1 & 13.75 & 0.69 & 4.86 & 0.07 & 19.3 & 2.3 & SB1\\
		 BS357$_\star$ & HD 240016 & 22:39:22.54 & 59:00:25.9 & 24.2 & 1.0 & 3.24 & 0.1 & 18.1 & 2.0 & 4.99 & 0.12 & 23.8 & 4.8 & SB1; R\\
		 BS358$_\star$ & TYC 4264-1036-1 & 22:34:06.50 & 60:58:50.2 & 20.5 & 0.7 & 3.83 & 0.08 & 4.27 & 0.16 & 3.42 & 0.05 & 6.8 & 0.52 & ...\\
		 BS361$_\star$ & TYC 4278-522-1 & 22:51:39.72 & 61:08:51.0 & 23.2 & 1.8 & 2.69 & 0.17 & 44.3 & 3.0 & 5.78 & 0.07 & 48.6 & 6.8 & SB1\\
		 BS362$_\star$ & ... & 22:55:45.4 & 60:24:40.3 & 25.1 & 2.1 & 3.34 & 0.21 & 15.2 & 1.1 & 5.02 & 0.11 & 59.4 & 13.0 & SB2\\
		 BS365$_\star$ & HD 2619 & 00:30:28.32 & 65:16:19.6 & 26.8 & 1.2 & 3.54 & 0.13 & 9.29 & 0.45 & 4.58 & 0.07 & 16.0 & 2.0 & ...\\
		 BS367$_\star$ & TYC 4046-1614-1 & 02:17:53.21 & 61:11:12.8 & 32.5 & 0.9 & 3.95 & 0.15 & 11.6 & 0.54 & 5.14 & 0.05 & 58.9 & 12.0 & ...\\
		 BS368$_\star$ & V* KM Cas & 02:29:30.46 & 61:29:44.2 & 32.3 & 1.0 & 3.91 & 0.17 & 9.26 & 0.5 & 4.93 & 0.06 & 53.4 & 29.0 & SB2\\
		 BS369$_\star$ & BD+60 586 & 02:54:10.68 & 60:39:03.6 & 37.3 & 1.0 & 4.22 & 0.1 & 11.18 & 1.2 & 5.34 & 0.10 & 68.0 & 21.0 & R\\
		 BS381$_\star$ & HD 54662 & 07:09:20.26 & -10:20:47.8 & 39.9 & 1.2 & 4.03 & 0.1 & 13.9 & 1.3 & 5.65 & 0.09 & 69.0 & 18.0 & ...\\
		 BS384$_\star$ & CD-35 4415 & 08:16:32.09 & -35:38:52.8 & 35.4 & 1.5 & 4.26 & 0.22 & 12.53 & 0.77 & 5.36 & 0.07 & 81.0 & 53.0 & SB2\\
		 BS386$_\star$ & V* GP Vel & 09:02:06.86 & -40:33:16.9 & 24.5 & 1.1 & 2.8 & 0.1 & 36.4 & 1.8 & 5.59 & 0.22 & 39.5 & 6.7 & EB\\
		 BS407$_\star$ & HD 93249 & 10:44:43.87 & -59:21:25.2 & 32.8 & 0.5 & 3.5 & 0.09 & 12.58 & 0.74 & 5.21 & 0.06 & 25.8 & 3.8 & ...\\
		 BS409$_\star$ & HD 93027 & 10:43:17.90 & -60:08:03.5 & 33.8 & 1.0 & 3.95 & 0.1 & 10.4 & 1.2 & 5.10 & 0.11 & 50.9 & 24.0 & ...\\
		 BS414$_\star$ & V* V738 Car & 10:48:46.55 & -60:35:40.5 & 26.7 & 1.0 & 3.13 & 0.1 & 11.41 & 0.58 & 4.67 & 0.07 & 15.7 & 1.5 & EB\\
		 BS634$_\star$ & HD 152756 & 16:57:15.00 & -43:43:15.2 & 31.7 & 0.7 & 4.18 & 0.1 & 11.2 & 0.51 & 5.06 & 0.05 & 65.9 & 16.0 & ...\\
		 BS640$_\star$ & HD 326533 & 16:58:24.05 & -42:43:41.2 & 27.3 & 1.5 & 4.28 & 0.16 & 10.13 & 0.65 & 4.74 & 0.09 & 71.0 & 23.0 & SB2; R\\
		 BS667$_\star$ & V* V1012 Sco & 17:15:22.32 & -38:12:46.8 & 29.6 & 0.7 & 4.05 & 0.1 & 10.05 & 1.1 & 4.85 & 0.09 & 55.8 & 31.0 & EB\\
		 BS673$_\star$ & ... & 17:20:38.23 & -38:01:48.7 & 27.1 & 1.8 & 4.01 & 0.29 & 9.1 & 1.8 & 4.57 & 0.21 & 14.4 & 1.5 & ...\\
		 BS687$_\star$ & ... & 17:28:59.78 & -36:06:54.7 & 27.1 & 2.1 & 4.19 & 0.35 & 19.6 & 9.9 & 5.28 & 0.48 & 78.0 & 59.0 & SB2\\
		 BS692$_\star$ & ... & 17:27:11.23 & -34:14:35.2 & 36.8 & 1.1 & 4.6 & 0.14 & 10.62 & 0.52 & 5.28 & 0.05 & 98.0 & 21.0 & SB2\\
		 BS700$_\star$ & ... & 17:33:47.88 & -31:16:27.1 & 27.0 & 1.9 & 3.82 & 0.25 & 25.0 & 11.0 & 5.62 & 0.35 & 84.0 & 25.0 & SB2\\
		 BS705$_\star$ & CD-29 13925 & 17:45:08.21 & -29:56:45.2 & 27.1 & 1.5 & 3.45 & 0.16 & 14.8 & 0.9 & 5.04 & 0.09 & 25.5 & 5.0 & ...\\
		 2G0063185-0066622$_\star$ & ... & 18:02:29.23 & -23:49:34.0 & 24.0 & 1.6 & 3.04 & 0.19 & 43.0 & 11.0 & 5.76 & 0.28 & 68.0 & 21.0 & SB2\\
		 2G0065094-0061344$_\star$ & ... & 18:02:41.88 & -23:38:02.0 & 27.9 & 1.7 & 3.71 & 0.18 & 7.35 & 0.49 & 4.45 & 0.10 & 15.1 & 2.5 & SB2\\
		 2G0118310-0062990$_\star$ & HD 166965 & 18:13:51.38 & -18:59:22.2 & 23.0 & 0.8 & 3.43 & 0.1 & 11.51 & 0.59 & 4.50 & 0.07 & 15.6 & 2.1 & ...\\
		 2G0592097+0012519$_\star$ & ... & 19:41:43.08 & 23:16:01.9 & 26.2 & 2.0 & 3.69 & 0.24 & 29.5 & 10.0 & 5.62 & 0.36 & 145.0 & 110.0 & SB2; R\\
		 2G0600615-0018340$_\star$ & HD 344878 & 19:44:44.06 & 23:51:07.9 & 29.8 & 0.7 & 3.96 & 0.12 & 8.18 & 0.37 & 4.69 & 0.05 & 20.3 & 4.1 & ...\\
		 2G0770694+0193568$_\star$ & HD 193427 & 20:18:45.17 & 39:24:22.3 & 30.1 & 0.6 & 4.03 & 0.09 & 7.24 & 0.55 & 4.60 & 0.07 & 19.2 & 3.8 & R\\
		 2G1045666+0128085$_\star$ & BD+57 2513 & 22:22:02.57 & 58:43:09.5 & 34.9 & 0.5 & 4.64 & 0.11 & 11.42 & 0.51 & 5.25 & 0.13 & 111.0 & 13.0 & SB2\\
		 $\kappa$ Cas & $\kappa$ Cas & 00:33:00.00 & 62:55:54.5 & 22.8 & 1.2 & 2.72 & 0.11 & 50.1 & 9.2 & 5.78 & 0.22 & 60.0 & 17.0 & ...\\
		 2G3525610+0001887$_\star$ & CD-35 11561 & 17:26:36.10 & -35:11:32.3 & 26.7 & 1.7 & 3.77 & 0.19 & 10.58 & 0.71 & 4.75 & 0.10 & 21.2 & 4.2 & SB2\\
		 2G3533026+0008095$_\star$ & HD 319881 & 17:28:21.67 & -34:32:30.5 & 39.6 & 1.8 & 3.97 & 0.14 & 13.86 & 0.73 & 5.64 & 0.07 & 67.0 & 18.0 & SB2\\
		 2G3585210+0089852$_\star$ & ... & 17:29:12.94 & -31:32:03.5 & 34.7 & 0.5 & 4.44 & 0.13 & 11.14 & 0.75 & 5.21 & 0.06 & 90.0 & 22.0 & SB2\\
	\enddata
	\tablecomments{(1) Stellar bow shock nebulae identifier from the K16 or addendum J19 catalogs; (2) Common alternative identifier, when available; (3) Right Ascension (J2000) in HH:MM:SS; (4) Declination (J2000) in DD:MM:SS; (5) Stellar effective temperature in kK; (6) Uncertainty in stellar effective temperature in kK; (7) Log surface gravity in cgs units; (8) Uncertainty in log surface gravity in cgs units; (9) Stellar radius in solar radii; (10) Uncertainty in stellar radius in solar radii; (11) log stellar luminosity in solar luminosity units; (12) uncertainty in log stellar luminoisty in solar luminosity units; (13) Stellar mass in solar masses; (14) Uncertainty in stellar mass in solar masses (15) binarity flag (from P25) with SB1 corresponding to a single-lined spectroscopic binary with significant radial velocity variations, SB2 indicating a double-line spectroscopic binary, R signifying a Gaia DR3 RUWE value $>1.4$, and EB indicating an eclipsing binary.}
\end{deluxetable*}
\end{longrotatetable}

Table~\ref{tab:BS_observed} lists the bow shock stars studied in this work. Column (1) presents the bow shock identifier from the K16 or the J19 catalogs. The star $\kappa$ Cas does not appear in the K16/J19 catalogs but was included in the sample of P25 because the central star is a blue supergiant and lies at the center of a bow shock nebula. Column (2) lists the common name for the bow shock star, when the bow shock star is a part of other catalogs. Columns (3) and (4) list the Right Ascension and Declination in J2000 coordinates. Columns (5) and (6) list the temperature and uncertainty in kK. Columns (7) and (8) list the log of the surface gravity and the uncertainty in cgs units. Columns (9) and (10) list the stellar radius and uncertainty in solar units. Columns (11) and (12) present the $\log$ of the stellar luminosity and the uncertainty in the $\log$ of stellar luminosity in solar luminosity units. Columns (13) and (14) list the stellar mass and uncertainty in solar units. Column (15) lists the binarity indicator from P25, with SB1 indicating a single-lined spectroscopic binary that has significant radial velocity variation, SB2 indicating a double-lined spectroscopic binary, R signifying a Gaia DR3 RUWE value $>1.4$, and EB indicating an eclipsing binary.

Our full list of targets spans galactic longitudes $\ell=0.1^\circ$--359.1$^\circ$ and galactic latitudes $b=-1.0^\circ$--$23.6^\circ$ with a median latitude of 2.9$^\circ$. Most bow shock stars are within the $\pm1^\circ$ of the galactic plane with the only outlier $\zeta$ Oph (BS013$_\star$) at a galactic latitude $+23.6^\circ$. The distances range from 135 pc for the closest star ($\zeta$ Oph, BS013$_\star$) to 11.5~kpc for the furthest star (BS303$_\star$) with a median distance of 2.02~kpc. The target stars span temperatures 20.0--46.8 kK, with a median temperature of 27.3~kK and $\logg$ 2.69--4.64, with a median $\logg$ of 3.83.

\section{Measuring $\dot{M}$}
\label{sec:measuringmdot}


In K18, the authors showed that Equation~\ref{eq:one} can be rewritten as:

\begin{equation}
\begin{split}
    &\dot{M}\left(\moperyear\right) =1.67\times10^{-28}\times\\
    &\frac{\left[R_0\left(\arcsecs\right)\right]^2D\left(\kpc\right)\left[V_a\left(\kms\right)\right]^2I_\nu\left(\Jysr\right)}{V_w\left(\kms\right)\ell\left(\arcsecs\right)j_\nu\left(\Jycmsrnuc\right)}\text{.}
\end{split}
    \label{eq:two}
\end{equation}







\noindent where $R_0$ is the standoff distance between the star and the apex of the nebula, $D$ is the distance to the bow shock star, $V_a$ is the velocity of the star with respect to its surroundings, $I_\nu$ is the specific intensity of the emitting dust in the bow shock nebula, $V_w$ is the terminal velocity of the stellar wind, $\ell$ is the chord length across the bow shock arc, and $j_\nu$ is the dust emissivity per nucleon of grains at the apex of the nebula. The expression is linear for the quantities $D$, $I_\nu$, $V_w$, $\ell$, and $j_\nu$, while $R_0$ and $V_a$ take a squared dependence.

\begin{figure*}[htb!]
        \plotone{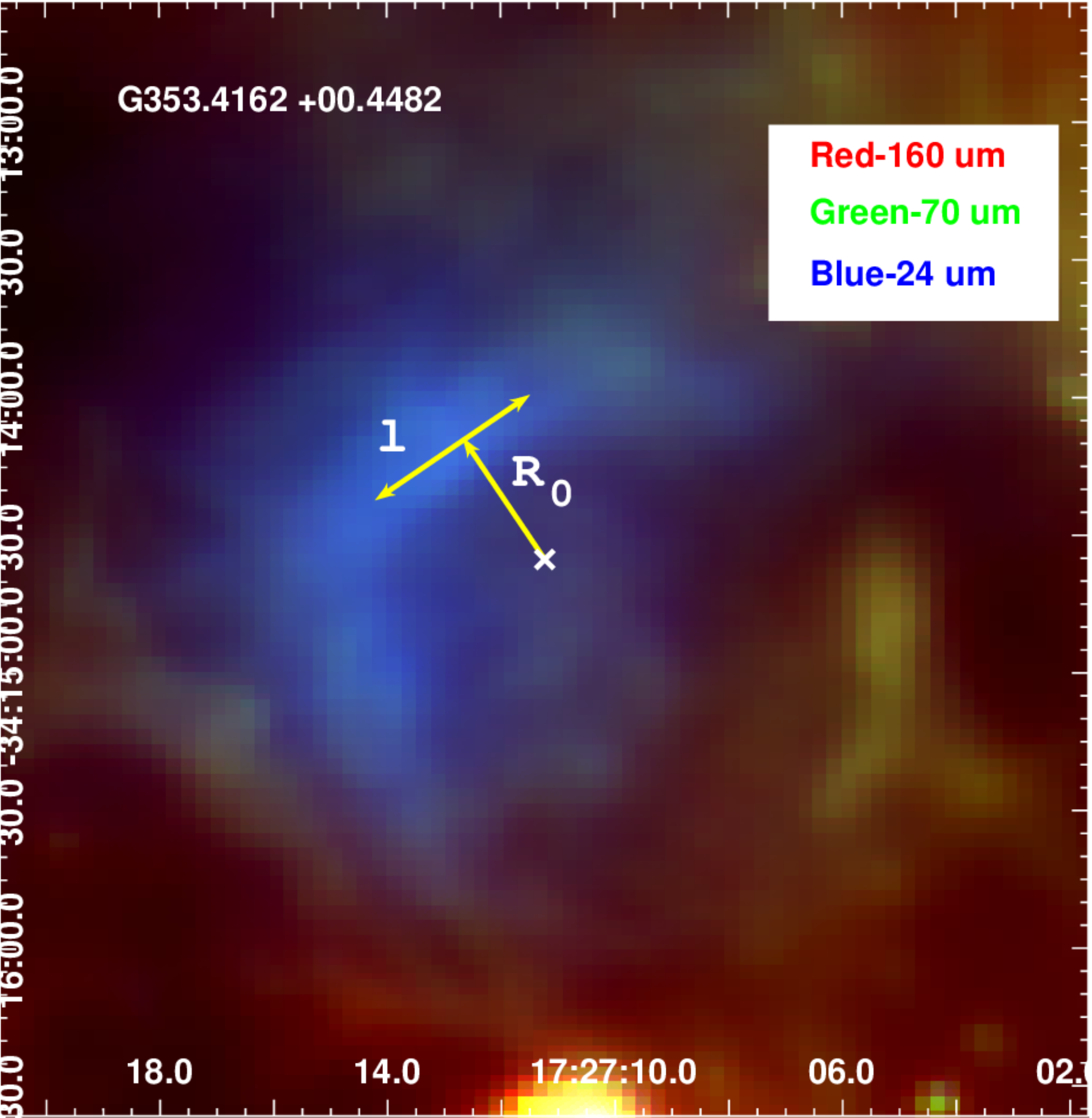}
        \caption{Three color image of object BS692$_\star$ from K16 with red/green/blue representing 160 and 70~$\um$ \textit{HSO} and 24 $\um$ \textit{SST} data respectively. The $\times$ marks the location of the driving star and the vectors depict the morphological parameters $R_0$ and $\ell$.}
        \label{fig:bowshock}
\end{figure*}


 
Figure~\ref{fig:bowshock} displays a three-color representation of the bow shock nebula surrounding G353.4162+00.4482 (bow shock candidate star BS692$_\star$ from K16). Red and green depict archival \textit{HSO} 160 and 70~$\um$ images and blue represents \textit{Spizer Space Telescope} (\textit{SST}) 24~$\um$ image data. We use the Multiband Imaging Photometer GALactic plane \citep{MIPSGAL} 24~$\um$ or Widefield Infrared Survey Explorer \citep[][WISE]{WISE} 22~$\um$ survey data throughout this work to assess the bow shocks' location and morphology. The bow shock is most evident at 24~$\um$ and is visible at 70~$\um$ but not at 160~$\um$. The bow shock is the dominant feature at 24~$\um$, but at 70~$\um$, both background and foreground nebulosity are visible. We obtain the chord length as the length across the bow shock arc perpendicular to the standoff distance vector. A white $\times$ marks the location of the driving OB star, which is not directly visible in the images. The yellow vectors illustrate the morphological parameters $R_0$ and $\ell$. To obtain $R_0$, we adopt the standoff distances measured by the citizen science project described in \citet{Whisnant2025}. They found overall good agreement between their crowd-sourced measurements for $R_0$ compared to single-measurement previous values from K16. For the subset of stellar bow shock nebulae with ``small'' standoff distances ($R_0\lesssim10''$) as measured in K16, \citet{Whisnant2025} reported their updated measurements of $R_0$ were systematically larger by $\sim$1\farcs5. We judge that the multiple measurements for $R_0$ obtained in \citet{Whisnant2025} are superior to the subjective measurements made by a single scientist in K16, so we adopt values from the former, when available. We take the measurements of the projected $R_0$ for unknown viewing angles to approximate the standoff distance at 0$^\circ$ since simulations have shown that the peak emission in the infrared is invariant for a wide range of viewing angles \citep{Acreman2016}.




Another required parameter is the velocity of the star relative to the surrounding ISM, $V_a$. K19 and K22 estimated this quantity using proper motion/parallax data from Gaia DR2 and EDR3, respectively. However, proper motions must be corrected for solar motion relative to the Local Stand of Rest (LSR) and the adoption of a Galactic circular velocity as a function of Galactocentric distance (\textit{i.e.}, a galactic rotation curve). K22 performed this analysis on bow shock stars with available proper motions and parallaxes as well as on a control sample of non-bow shock GOSC stars. \citet[][hereafter C23]{Carretero-Castrillo23} performed a similar runaway analysis of Galactic O stars using Gaia DR3 proper-motion data with some differences in their methodology compared to K22. C23 applied the zero-point correction to the Gaia parallaxes \citep[described in][]{Lindegren2021} while K22 did not. C23 also used a different rotation curve and solar motion with respect to the LSR to compute the peculiar 2D space velocity.

\begin{figure}[htb!]
        \plotone{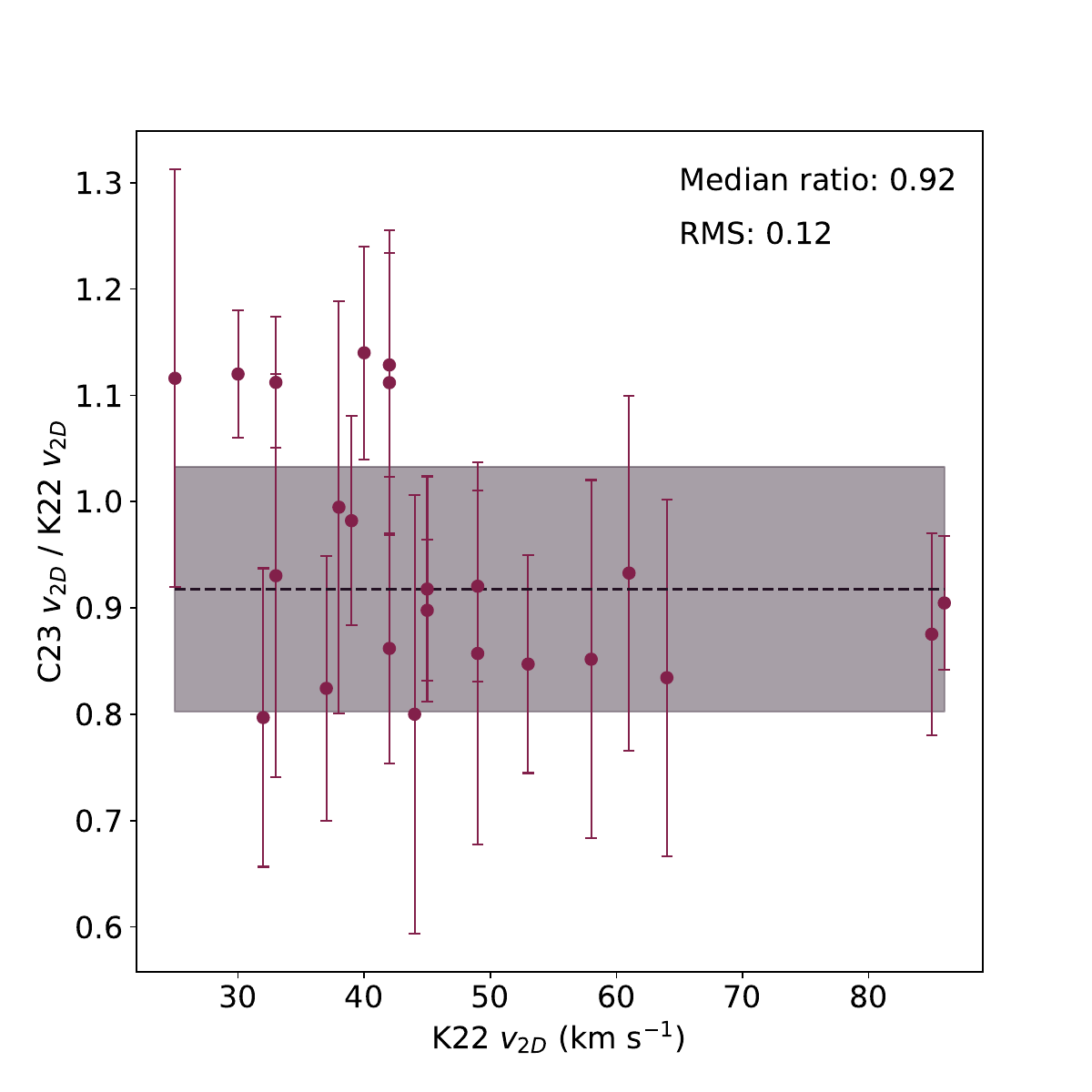}
        \caption{The ratio of the C23 to the K22 peculiar space velocities as a function of K22 peculiar space velocity for the 23 GOSC stars studied in both works. The dashed line and the shaded region represent the median and RMS of the ratio.}
        \label{fig:GOSC_comparison}
\end{figure}



Figure~\ref{fig:GOSC_comparison} plots the ratio of the C23 to the K22 peculiar 2D space velocities for the overlapping subsample of 23 GOSC stars analyzed in both works as a function of the K22 values. The horizontal dashed black line shows the median ratio of 0.92. The grey band encompasses the RMS of 0.12. Overall, there is good agreement between the two independent determinations of velocity. Because of a difference in methodology, C23 reported slightly higher uncertainties than K22. The error bars in Figure~\ref{fig:GOSC_comparison} represent the K22 and C23 uncertainties added in quadrature. Low-velocity stars on the left side of the plot have a typical uncertainty of 5~$\kms$, which corresponds to 20\%, while the fastest stars at 90~$\kms$ have the same nominal uncertainty, amounting to only 7\%. The differences between K22 and C23 do not correlate with galactic longitude as might be expected if systematic differences in the galactic rotation curve were responsible. The difference between the K22 and C23 values could arise from systematics such as the choice of mean solar motion with respect to the LSR or the application of the zero-point offset correction. Nevertheless, the K22 and C23 velocites correlate with each other and are in agreement within the reported uncertainties.







In Equation~\ref{eq:two}, we take $V_W$ to be the terminal wind speed of the stellar outflows $v_\infty$. The terminal wind speed is measurable from the P-Cygni profiles of UV metal-resonance lines, which all OB stars exhibit. \citet{Mokiem2007} tabulated measurements from UV-based spectroscopy of 49 Galactic O- and early B-type stars with effective temperatures ranging from 25--47~kK (Table A.1 in their Appendix). They reported terminal wind speeds ranging from 1500~$\kms$ for the coolest stars to 3200~$\kms$ for the hottest stars. Since UV spectra do not exist for our sample of bow shock stars\footnote{Exceptions $\zeta$ Oph (BS013$_\star$), HD 195592 (BS343$_\star$), HD 199021 (BS345$_\star$), HD 41161 (BS$374_\star$), V* AE Aur (BS375$_\star$), HD 48099 (BS375$_\star$), $\upsilon$ Ori (BS378$_\star$), HD 54662 (BS381$_\star$), V* GP Vul (BS386$_\star$), HD 153426 (BS662$_\star$), V* V1012 Sco (BS667$_\star$), and $\kappa$ Cas (BS794$_\star$).}, K18/K19 used the results of \citet{Mokiem2007} to estimate the terminal wind speed for their sample, with $\teff$ approximated from rough spectral types in the literature. In this work, we adopt the $v_\infty$--$\teff$ calibration  of \citet{Hawcroft2024}, which used 66 Galactic OB stars spanning 15 kK -- 45 kK from the UV spectroscopic study of \citet{HowarthandPrinja1989} with updated stellar parameters provided by \citet{Holgado2020}. The \citet{Hawcroft2024} relation shows little-to-no dependence on luminosity class---no $v_\infty$ dependence on $\logg$. Their relation shows an RMS of 277~$\kms$. We add this RMS in quadrature to the error produced by the uncertainty in temperature measured in P25.

For distance $D$ we adopt the inverse parallactic distance provided by Gaia DR3 without any zero-point correction given the large distances to these stars.

We measure the specific intensity at 70~$\um$ $I_\nu$ from archival \textit{HSO} 70~$\um$ UNIMAPB level 2.5 images from the Photodetector Array Camera and Spectrometer (PACS) photometer. In cases where there is structure in the bow shock or confounding background emission, we find it necessary to smooth the images by 1--3 pixels to increase the signal-to-noise ratio. In most cases, the peak specific intensity at 70~$\um$ coincides with the peak specific intensity at 24~$\um$. However in some instances where the field has contaminating background and foreground emission, the peak location at 70~$\um$ is unclear. In such cases, we assume the two peaks are coincident and use the peak location at 24~$\um$ to measure the peak at 70~$\um$\footnote{It is possible that the two peaks may not be coincident if the 70~$\um$ band is dominated by larger dust grains which may be exterior to the shock front where the smaller grains are located.}. Using adjacent regions on either side of the peak along the line of $R_0$, we estimate the background flux by fitting a linear baseline. We subtract this estimate of the background from the peak brightness to measure the background-subtracted specific intensity $I_\nu$. We adopt an uncertainty on $I_\nu$ of 20\% based on the typical fluctuations in the background signal.

\begin{figure*}[htb!]
        \plottwo{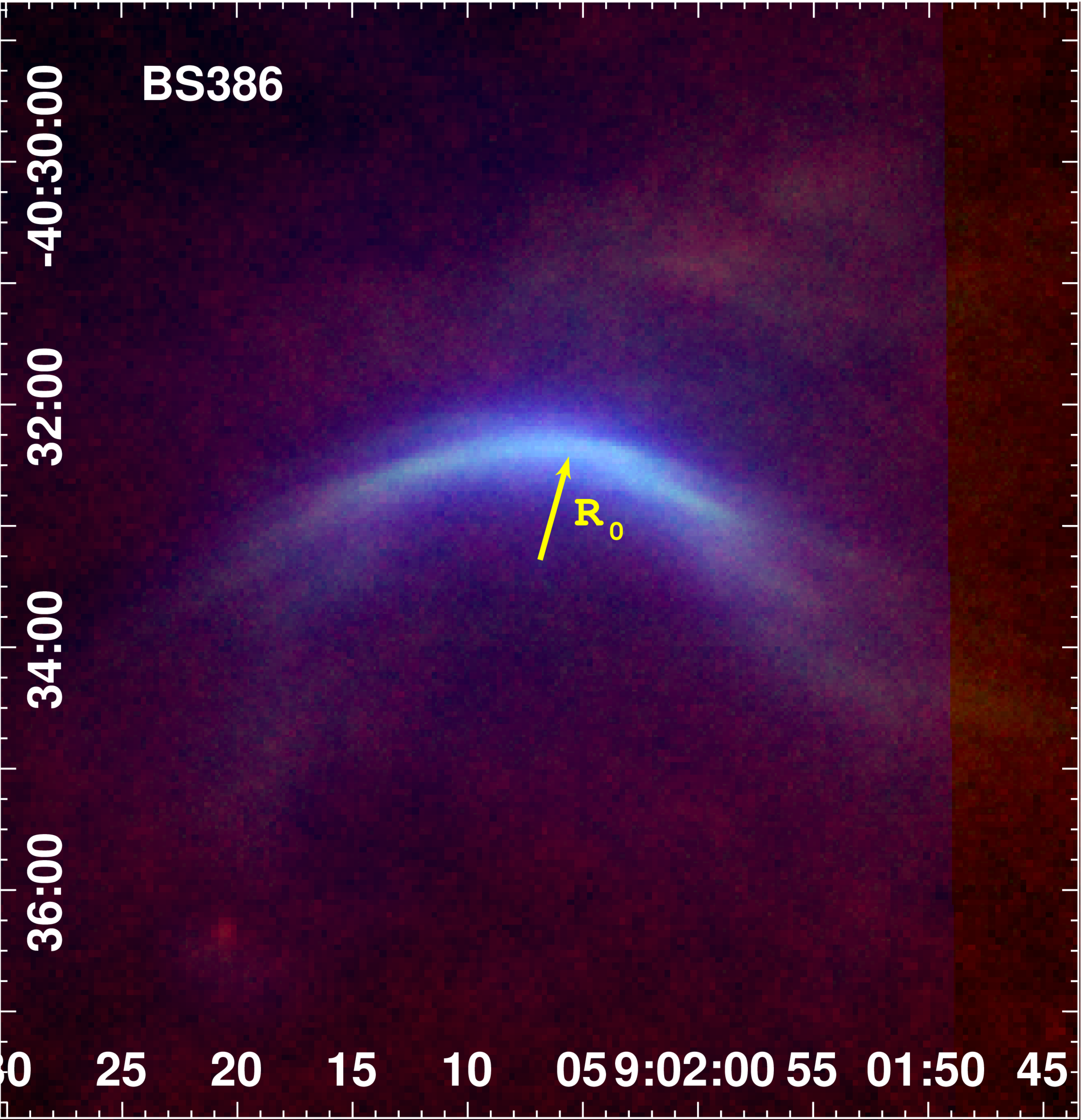}{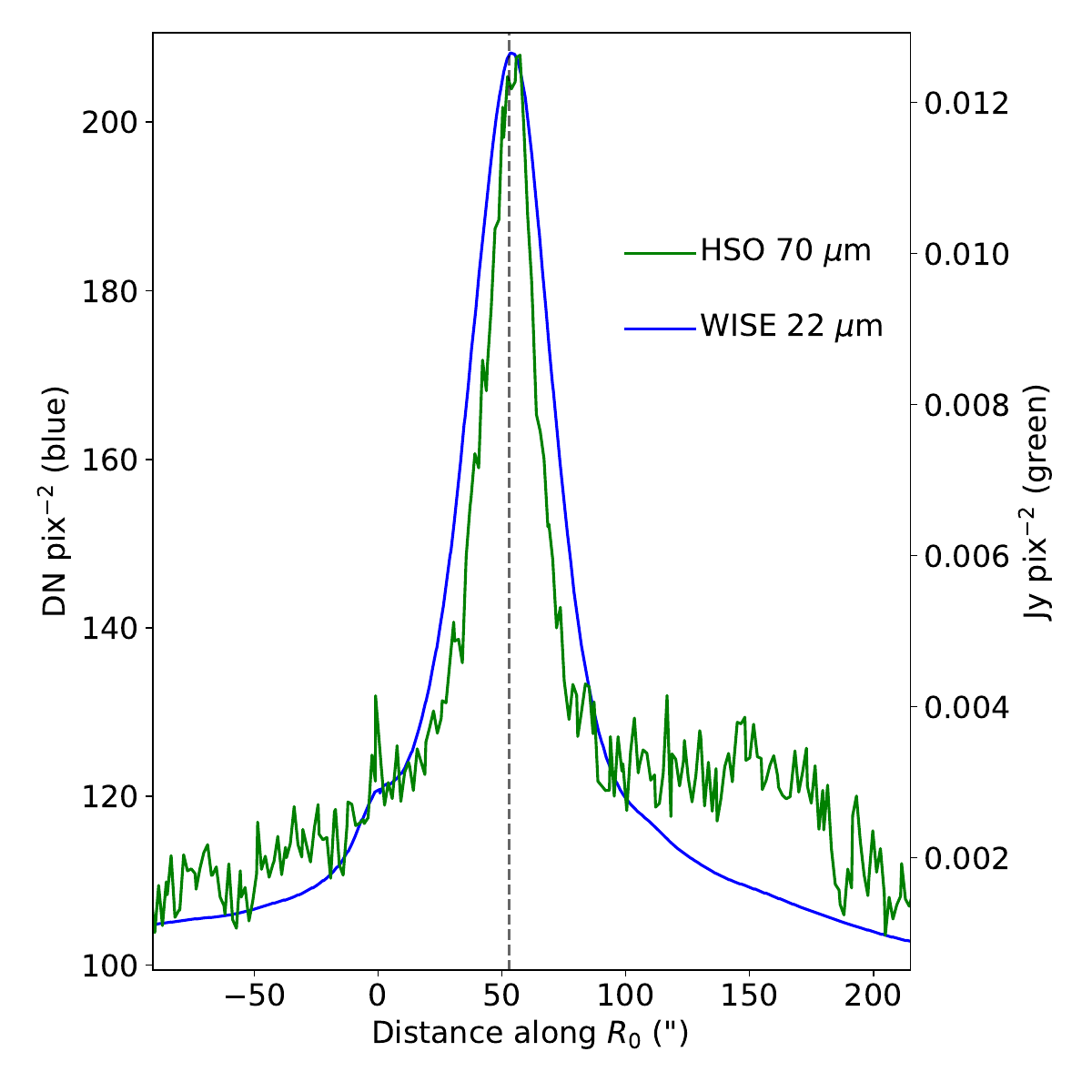}
        \caption{Left: Three color image of bow shock BS386 (surrounding the stellar X-ray source Vela X-1) with red/green/blue representing 160 and 70~$\um$ \textit{HSO} and 22~$\um$ \textit{WISE} data, respectively. Right: The brightness of the nebula along $R_0$, relative to the stellar source at 0, in 22~$\um$ (blue) and 70~$\um$ (green).}
        \label{fig:bowshock_multi}
\end{figure*}

Figure~\ref{fig:bowshock_multi} (left panel) displays a three-color representation of the bow shock nebula surrounding BS386$_\star$ in K16 in the same color mapping as in Figure~\ref{fig:bowshock}. The right panel displays the specific intensity of the bow shock nebula along the $R_0$ vector, relative to the stellar source at 0, with the left axis corresponding to the 22~$\um$ data in units of \texttt{W4} DN and the right axis corresponding to the 70~$\um$ data in units Jy pix$^{-2}$. The 22~$\um$ peak brightness is 208~DN, and the 70~$\um$ peak brightness is $1.25\times10^{-2}$~$\Jypersquarepix$ for this nebula. The peak at 70~$\um$ is coincident with the peak at 22~$\um$.








Most of the parameters in Equation~\ref{eq:two} are readily measurable from optical and infrared data and require few assumptions. The major exception to this is $j_\nu$, which must rely on an adopted set of theoretical properties for dust grains. We adopt the \citet[][D21 hereafter]{Draine2021} models to compute $j_\nu$. They modeled the infrared dust emission as a function of different radiation field intensities $U$, heating stellar populations, and dust compositions. $U$ is defined as

\begin{equation}
    U = \gamma_*\frac{u_*}{u_{\text{mMMP}}}\text{ ,}
    \label{eq:three}
\end{equation}

\noindent where $\gamma_\star$ is a small model-dependent factor of order 1--5, and $u_{\text{mMMP}}$ is the modified mean interstellar radiation field (ISRF) used by D21 ($u_{\text{mMMP83}}=3.1293\times10^{-2}~\ergscmc$), which in turn is based on the mean ISRF energy density calculated by \citet[][$u_{\text{MMP83}}=2.17\times10^{-2}~\ergscmc$]{MMP83}. We compute $u_*$---the energy density at the apex of the infrared nebula---from

\begin{equation}
    u_* = \frac{R_*^2\sigma\teff^4}{R_0^2c}\text{ ,}
    \label{eq:four}
\end{equation}

\noindent where $R_*$ is the radius of the star, $\teff$ is the effective temperature of the star, and $R_0$ is the standoff distance. D21 calculated the infrared emission spectra for three different irradiating spectral energy distributions: young stellar populations of different ages, an updated ISRF, and M31 bulge stars. They used the BPASS \citep{BPASS, BPASS2} and BC03 \citep{BC03} stellar evolution models to provide the input spectral energy distributions for the young star populations. They considered several different dust compositions that included different PAH ionization distributions (high, standard, and low) and particle size distributions (large, standard, and small). We adopt their emissivities calculated from the BPASS models at solar metallicity and age 10~Myr to best reflect the stars at the center of bow shock nebulae. We also select the models with standard grain size and ionization distributions. We interpolate between models to obtain $j_\nu$ as a function of $U$. Our approach differs from the one adopted in K19 in that we use the model $j_\nu$ created specifically for young stellar populations, so we do not require the ad-hoc correction suggested by \citet{Henney2019b}.


\begin{figure}[htb!]
        \plotone{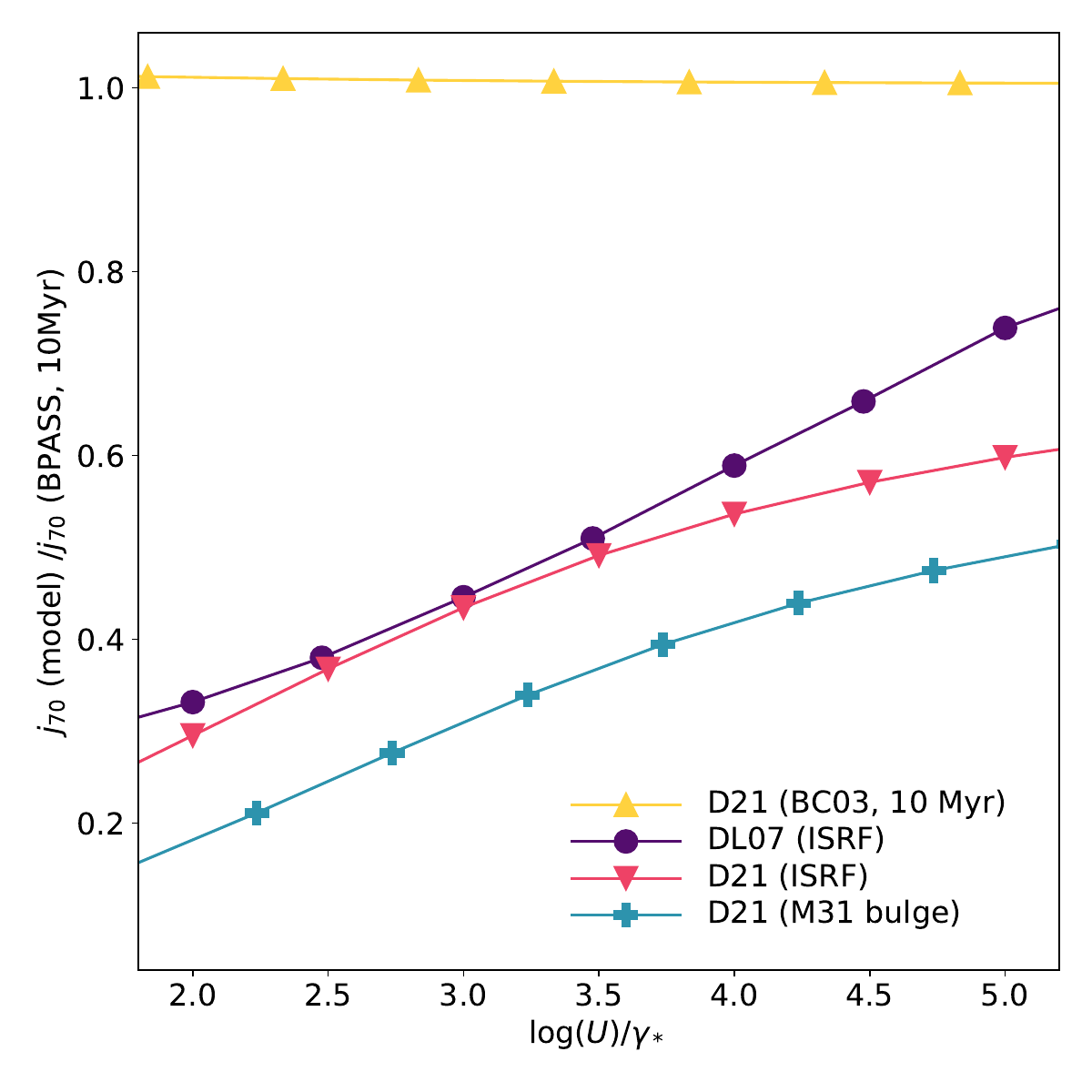}
        \caption{Normalized dust emissivities at 70~$\um$ for several different D21 and DL07 models as a function of irradiation parameter $U/\gamma_\star$.}
        \label{fig:compare_DL07}
\end{figure}




We consider whether the choice of model makes a difference on dust emissivity $j_\nu$ at 70~$\um$ by plotting in Figure~\ref{fig:compare_DL07} a normalized $j_\nu$ versus $U/\gamma_\star$ for a subset set of dust emission models from D21 and \citet[][DL07 hereafter]{DraineandLi2007}. Each set of models is normalized to the dust emissivity from the D21 BPASS model at age 10 Myr and solar metallicity. The figure shows the effects of different heating efficiencies of dust grains for different age stellar populations. The yellow track with filled triangles denotes the D21 BC03 model at 10 Myr and solar metallicity. It lies near unity at every irradiation intensity, indicating excellent agreement between the D21 BPASS and D21 BC03 models at the same age and metallicity. The other three models (DL07 ISRF---purple circles, D21 ISRF---pink up side down triangles, and D21 M31 bulge---cyan $+$'s) have ratios substantially less than 1.0 for some regimes (with ratios ranging from 0.2--0.8) showing emissivities of grains heated by old stellar populations depart significantly from emissivities of grains heated by young stellar populations. The DL07 ISRF and D21 ISRF models generally show good agreement with each other and are factors of 1.5--3 lower than the D21 BPASS model, with larger deviations on the lower $U/\gamma_\star$ end. This factor of 1.5--3 difference represents the ad-hoc correction K19 applied to the DL07 ISRF dust emissivities. We expect that the D21 BPASS model---tailored specifically for irradiation by young stellar populations---to better reflect the infrared dust properties of bow shock nebulae around early type stars. Accordingly, we adopt the D21 BPASS model emissivities as most suitable for use in this work.




\begin{figure}[htb!]
        \plotone{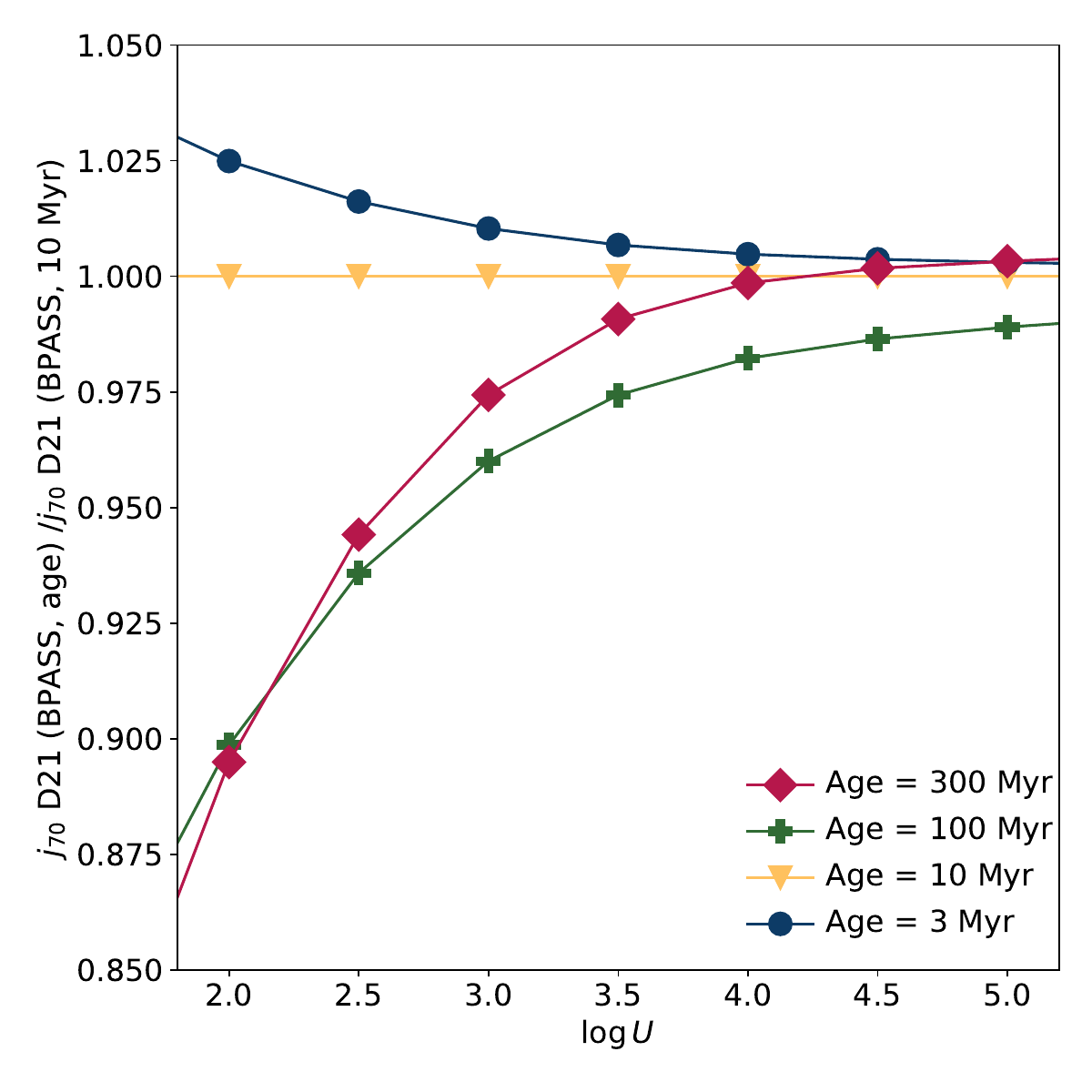}
        \caption{Normalized dust emissivities at 70 $\um$ for the D21 BC03 models at solar metallicity versus irradiation parameter $U$ for different stellar population ages.}
        \label{fig:compare_ages}
\end{figure}

We also consider differences that could arise in our choice of ages for the irradiating stellar populations. Figure~\ref{fig:compare_ages} plots the dust emissivity calculated at 70~$\um$ heated by single-age D21 BPASS models, for isochrone ages 3, 10, 100 and 300 Myr, versus radiation intensity $U$. The calculated dust emissivities are normalized to the 10 Myr model. All emissivities agree to within $\sim$5\% over the range of $3\leq \log U\leq5$. Relatively larger deviations are exhibited for the 100 and 300 Myr models on the low-$U$ end, with differences exceeding 12\%. We find that while the choice of age matters in obtaining $j_\nu$, differences are negligible between the 3 and 10 Myr models. Since the central stars to bow shock nebulae are almost exclusively OB stars \citep[][P25]{Chicketal2020}, the 3 and 10 Myr comparisons are most pertinent to this study, so we adopt the 10 Myr model. The other model free parameters---grain size and ionization distributions---affect the dust emissivity to much smaller degrees.



\begin{longrotatetable}
\begin{deluxetable*}{lccccccccccccccc}
	\tablecaption{Mass-loss parameters for bow shock stars}\label{tab:BSstars}
	\tabletypesize{\scriptsize}
	\tablehead{
		ID & $R_0$ & $\sigma_{R_{0}}$ & $D$ & $\sigma_D$ & $V_{2D}$ &$\sigma_{V_{2D}}$ & $I_\nu$ & $\log U$ & $V_w$ & $\ell$ & $\log j_\nu$ & $n_a$ & Morph. & PAH & Conf.\\& $(\arcsec)$ & $(\arcsec)$ & (pc) & (pc) & $(\kms)$ & $(\kms)$ &(MJy sr$^{-1})$ & (dex )& $(\kms)$ & $(\arcsec)$ & $(\text{Jy cm}^2\text{ sr}^{-1}$ & $(\text{cm}^{-3})$ & && \\&&&&&&&&&&&$\text{nucleon}^{-1}$)&&&&\\(1) & (2) & (3) & (4) & (5) & (6) & (7) & (8) & (9) & (10) & (11) & (12) & (13) & (14) & (15) & (16)}
	\startdata
		BS001 & 27.77 & 4.61 & 2851 &144 & 19 & 2 & 615 & 3.66 & 2174 & 46.10 & -11.19 & 12.20 & Y & Y & H\\
		BS007 & 11.13 & 1.57 & 1217 &25 & 11 & 1 & 1660 & 3.49 & 976 & 11.20 & -11.29 & 392.00 & Y & Y & L\\
		BS010 & 3.10 & 0.31 & 2462 &101 & 19 & 2 & 81 & 5.07 & 1538 & 41.80 & -10.56 & 0.48 & N & N & L\\
		BS013 & 299.00 & 29.90 & 135 &12 & 23 & 3 & 97 & 3.92 & 1979 & 352.00 & -11.06 & 3.90 & Y & Y & H\\
		BS016 & 26.54 & 1.09 & 3069 &186 & 5 & 2 & 126 & 3.63 & 1496 & 20.70 & -11.21 & 5.33 & Y & N & H\\
		BS019 & 11.66 & 0.89 & 3198 &174 & 26 & 2 & 299 & 4.65 & 1199 & 12.80 & -10.73 & 6.55 & Y & N & H\\
		BS026 & 8.60 & 0.76 & 3270 &154 & 8 & 2 & 189 & 4.54 & 1606 & 9.35 & -10.78 & 6.15 & Y & N & H\\
		BS027 & 7.54 & 0.90 & 1802 &49 & 6 & 2 & 213 & 3.87 & 998 & 6.68 & -11.08 & 35.90 & Y & N & L\\
		BS043 & 6.80 & 0.68 & 2248 &222 & 6 & 2 & 61 & 4.44 & 1444 & 8.82 & -10.82 & 3.38 & Y & N & H\\
		BS051 & 2.50 & 0.25 & 1418 &45 & 3 & 1 & 5280 & 6.36 & 2172 & 2.80 & -10.11 & 283.00 & Y & Y & L\\
		BS056 & 28.00 & 2.80 & 5144 &984 & 32 & 32 & 28 & 3.56 & 1429 & 17.20 & -11.24 & 0.93 & Y & N & H\\
		BS061 & 22.44 & 1.55 & 2078 &157 & 10 & 2 & 769 & 4.23 & 1820 & 17.70 & -10.91 & 28.70 & N & N & L\\
		BS063 & 3.50 & 0.35 & 1758 &56 & 5 & 1 & 2280 & 5.34 & 1870 & 36.70 & -10.46 & 16.90 & Y & Y & L\\
		BS065 & 6.10 & 2.28 & 1814 &56 & 3 & 1 & 529 & 4.06 & 862 & 4.69 & -10.99 & 102.00 & Y & N & L\\
		BS067 & 8.47 & 0.99 & 1861 &66 & 4 & 1 & 2090 & 4.75 & 1984 & 8.14 & -10.69 & 112.00 & Y & Y & L\\
		BS150 & 12.10 & 1.78 & 613 &10 & 8 & 2 & 810 & 4.96 & 1263 & 37.90 & -10.60 & 23.30 & N & Y & L\\
		BS200 & 10.13 & 0.94 & 4523 &452 & 34 & 13 & 100 & 4.56 & 1264 & 11.10 & -10.77 & 1.94 & Y & N & H\\
		BS255 & 2.00 & 0.20 & 1272 &25 & 19 & 2 & 149 & 5.23 & 741 & 4.11 & -10.50 & 15.10 & N & N & L\\
		BS288 & 11.76 & 1.06 & 2911 &108 & 11 & 2 & 72 & 3.81 & 1478 & 29.50 & -11.11 & 1.82 & Y & N & H\\
		BS300 & 6.90 & 0.69 & 2053 &47 & 4 & 1 & 43 & 4.45 & 1675 & 32.30 & -10.81 & 0.71 & Y & N & H\\
		BS302 & 4.90 & 0.49 & 2452 &159 & 7 & 2 & 37 & 4.32 & 940 & 24.40 & -10.87 & 0.76 & Y & N & H\\
		BS303 & 7.80 & 0.78 & 11455 &2441 & 52 & 52 & 54 & 4.24 & 763 & 7.61 & -10.91 & 0.84 & N & N & L\\
		BS305 & 4.90 & 0.49 & 1922 &329 & 9 & 5 & 582 & 5.42 & 2119 & 21.70 & -10.43 & 6.24 & N & N & L\\
		BS306 & 13.67 & 3.41 & 2180 &63 & 7 & 1 & 48 & 4.01 & 1479 & 34.70 & -11.02 & 1.10 & Y & N & H\\
		BS314 & 5.40 & 0.54 & 4525 &297 & 11 & 2 & 59 & 4.36 & 1408 & 20.00 & -10.86 & 0.78 & Y & N & H\\
		BS320 & 35.50 & 3.55 & 3179 &166 & 7 & 2 & 81 & 3.68 & 1531 & 30.90 & -11.18 & 2.11 & N & Y & L\\
		BS321 & 10.50 & 1.05 & 4748 &253 & 1 & 1 & 11 & 3.76 & 1854 & 12.70 & -11.14 & 0.43 & N & Y & L\\
		BS322 & 19.50 & 1.95 & 4916 &249 & 3 & 1 & 50 & 3.64 & 2031 & 20.30 & -11.21 & 1.35 & Y & N & H\\
		BS324 & 25.70 & 2.57 & 5325 &303 & 30 & 2 & 27 & 3.50 & 2001 & 34.90 & -11.28 & 0.46 & Y & N & H\\
		BS325 & 12.90 & 1.29 & 1318 &38 & 12 & 2 & 24 & 4.50 & 900 & 16.30 & -10.79 & 1.14 & Y & N & H\\
		BS330 & 25.50 & 2.55 & 1711 &52 & 9 & 1 & 121 & 4.28 & 985 & 53.00 & -10.89 & 1.73 & Y & N & H\\
		BS331 & 45.60 & 4.56 & 1691 &39 & 4 & 2 & 98 & 4.10 & 2243 & 35.00 & -10.97 & 2.61 & Y & N & H\\
		BS332 & 28.20 & 2.82 & 1732 &42 & 5 & 1 & 1220 & 3.78 & 1981 & 24.80 & -11.13 & 64.30 & N & N & L\\
		BS335 & 15.60 & 1.56 & 2012 &49 & 12 & 2 & 138 & 4.27 & 1184 & 18.80 & -10.90 & 4.80 & N & N & H\\
		BS339 & 70.00 & 7.00 & 1582 &45 & 77 & 3 & 384 & 4.21 & 3353 & 54.30 & -10.92 & 6.22 & N & Y & L\\
		BS340 & 15.90 & 1.59 & 1473 &45 & 19 & 2 & 41 & 4.26 & 1337 & 37.60 & -10.90 & 0.97 & Y & Y & L\\
		BS341 & 20.20 & 2.02 & 1770 &48 & 11 & 2 & 90 & 4.48 & 2076 & 25.20 & -10.80 & 2.13 & Y & N & H\\
		BS344 & 193.00 & 19.30 & 1717 &36 & 45 & 2 & 698 & 3.56 & 3470 & 193.00 & -11.24 & 6.18 & Y & Y & H\\
		BS350 & 23.00 & 2.30 & 2316 &433 & 10 & 2 & 27 & 3.77 & 1170 & 45.10 & -11.14 & 0.60 & Y & N & H\\
		BS353 & 30.50 & 3.05 & 4645 &233 & 67 & 4 & 34 & 3.60 & 2545 & 26.90 & -11.23 & 0.77 & Y & N & H\\
		BS356 & 15.60 & 1.56 & 2791 &97 & 10 & 2 & 22 & 3.93 & 1375 & 13.60 & -11.06 & 1.10 & Y & N & H\\
		BS357 & 21.80 & 2.18 & 2738 &272 & 8 & 2 & 24 & 3.75 & 1166 & 33.60 & -11.15 & 0.62 & Y & N & H\\
		BS358 & 8.90 & 0.89 & 868 &9 & 2 & 1 & 20 & 3.99 & 794 & 10.80 & -11.03 & 3.82 & Y & N & H\\
		BS361 & 91.70 & 9.17 & 4564 &267 & 29 & 2 & 19 & 2.76 & 1063 & 80.40 & -11.74 & 0.48 & Y & N & H\\
		BS362 & 14.70 & 1.47 & 3894 &174 & 27 & 2 & 33 & 3.70 & 1256 & 20.80 & -11.17 & 1.00 & Y & Y & H\\
		BS365 & 29.90 & 2.99 & 1058 &19 & 11 & 2 & 23 & 3.90 & 1439 & 37.00 & -11.07 & 1.15 & Y & Y & H\\
		BS367 & 36.00 & 3.60 & 2100 &59 & 8 & 2 & 55 & 3.67 & 2013 & 40.20 & -11.19 & 1.67 & Y & Y & H\\
		BS368 & 14.00 & 1.40 & 1981 &62 & 8 & 1 & 199 & 4.34 & 1999 & 32.30 & -10.86 & 3.81 & Y & Y & H\\
		BS369 & 73.00 & 7.30 & 1961 &193 & 9 & 3 & 50 & 3.33 & 2505 & 72.20 & -11.38 & 1.43 & Y & Y & H\\
		BS381 & 236.00 & 23.60 & 1382 &100 & 11 & 3 & 38 & 2.92 & 2769 & 141.00 & -11.63 & 1.41 & Y & N & H\\
		BS384 & 31.00 & 3.10 & 3421 &142 & 4 & 2 & 228 & 3.59 & 2308 & 26.10 & -11.23 & 7.24 & Y & Y & H\\
		BS386 & 53.00 & 5.30 & 2015 &62 & 48 & 2 & 163 & 3.88 & 1202 & 89.70 & -11.08 & 1.82 & Y & Y & H\\
		BS407 & 7.80 & 0.78 & 2521 &112 & 10 & 2 & 1890 & 4.93 & 2046 & 18.80 & -10.62 & 27.60 & Y & Y & H\\
		BS409 & 7.40 & 0.74 & 2921 &206 & 19 & 2 & 778 & 4.73 & 2148 & 13.30 & -10.70 & 16.60 & Y & Y & H\\
		BS414 & 5.80 & 0.58 & 2350 &81 & 7 & 1 & 108 & 4.80 & 1423 & 10.30 & -10.67 & 3.47 & Y & Y & H\\
		BS634 & 32.61 & 1.71 & 1604 &45 & 9 & 2 & 75 & 3.92 & 1934 & 41.70 & -11.06 & 2.15 & Y & N & H\\
		BS640 & 6.95 & 0.68 & 1561 &65 & 12 & 2 & 499 & 4.94 & 1486 & 20.80 & -10.61 & 10.50 & Y & N & H\\
		BS667 & 16.47 & 1.43 & 1003 &36 & 29 & 2 & 643 & 4.71 & 1724 & 28.50 & -10.71 & 19.10 & Y & N & H\\
		BS673 & 6.00 & 0.60 & 3882 &307 & 21 & 3 & 150 & 4.17 & 1460 & 7.07 & -10.94 & 7.99 & Y & N & H\\
		BS687 & 9.29 & 0.71 & 2693 &152 & 20 & 3 & 131 & 4.77 & 1467 & 8.85 & -10.68 & 4.37 & Y & N & H\\
		BS692 & 32.08 & 2.09 & 1794 &62 & 17 & 1 & 1510 & 4.05 & 2449 & 52.70 & -11.00 & 26.70 & Y & Y & H\\
		BS700 & 15.80 & 1.58 & 3868 &310 & 16 & 3 & 66 & 4.20 & 1458 & 13.80 & -10.93 & 1.74 & Y & N & H\\
		BS705 & 26.52 & 1.04 & 2349 &100 & 24 & 3 & 1570 & 3.74 & 1465 & 26.40 & -11.15 & 60.10 & Y & N & H\\
		2G0063185-0066622 & 34.50 & 3.45 & 2705 &221 & 8 & 2 & 1060 & 4.10 & 1146 & 24.90 & -10.97 & 24.70 & Y & Y & H\\
		2G0065094-0061344 & 11.95 & 1.16 & 1983 &92 & 5 & 3 & 698 & 4.02 & 1548 & 10.30 & -11.01 & 58.70 & Y & Y & H\\
		2G0118310-0062990 & 9.70 & 0.97 & 1820 &67 & 5 & 2 & 95 & 4.33 & 1045 & 7.15 & -10.87 & 9.00 & Y & N & H\\
		2G0592097+0012519 & 17.65 & 3.72 & 1708 &100 & 4 & 2 & 90 & 4.91 & 1375 & 15.20 & -10.62 & 2.44 & N & N & L\\
		2G0600615-0018340 & 20.13 & 2.02 & 2090 &58 & 8 & 1 & 37 & 3.73 & 1744 & 17.90 & -11.16 & 2.37 & Y & N & H\\
		2G0770694+0193568 & 19.60 & 1.96 & 1429 &95 & 4 & 2 & 62 & 3.99 & 1770 & 39.90 & -11.03 & 1.93 & Y & N & H\\
		2G1045666+0128085 & 16.30 & 1.63 & 2238 &88 & 4 & 2 & 126 & 4.42 & 2261 & 29.40 & -10.83 & 2.16 & N & Y & L\\
		$\kappa$ Cas & 158.00 & 15.80 & 939 &144 & 23 & 5 & 35 & 3.74 & 1028 & 115.00 & -11.15 & 0.76 & Y & Y & H\\
		2G3525610+0001887 & 13.51 & 1.41 & 1855 &66 & 16 & 3 & 40 & 4.21 & 1424 & 29.50 & -10.92 & 1.03 & Y & N & H\\
		2G3533026+0008095 & 56.63 & 4.90 & 1792 &56 & 24 & 1 & 56 & 3.91 & 2736 & 77.20 & -11.06 & 0.78 & N & Y & L\\
	\enddata
	\tablecomments{(1) Stellar bow shock nebulae identifier from the K16/J19 catalogs; (2) Projected stand-off distance from \citet{Whisnant2025} and K16; (3) The uncertainty in the measured stand-off distance; (4) The parallactic distance to the bow shock central star; (5) The propagated uncertainty in the parallactic distance; (6) The peculiar 2D space velocity of the bowhsock central stars from K22; (7) The uncertainty in the computed peculiar 2D space velocity; (8) The background-subtracted specific intensity of the bow shock arc at 70~$\um$; (9) The logarithm of the irradiation parameter computed from Equations~\ref{eq:three} and \ref{eq:four}; (10) The interpolated terminal wind speed from the \citet{Hawcroft2024} relation; (11) The chord length across the bow shock arc; (12) The calculated dust-emissivity per nucleon of the dust grains at the apex of the bow shock arc; (13) The calculated interstellar density from Equation 5 of K18; (14) bow shock morphology indicator; (15) PAH indicator; (16) Confidence flag, assessed from the flags in Columns (13), (14), and (15)}
\end{deluxetable*}
\end{longrotatetable}


Table~\ref{tab:BSstars} presents all of the measured parameters needed to measure mass-loss rate $\mdot$. Column (1) presents the bow shock identifier from the K16/J19 catalogs. Columns (2) and (3) list the projected standoff distance $\rnaught$ and its uncertainty in arcseconds, sourced from \citet{Whisnant2025}, when available, or K16. \citet{Whisnant2025} provided the uncertainty in their citizen-science measurements on $\rnaught$ which we list in Column (3). For bow shocks with no citizen-science $\rnaught$ measurements, we assume a 10\% uncertainty. Columns (4) and (5) present the Gaia DR3 parallactic distance and uncertainty in pc. Columns (6) and (7) present the peculiar 2D space velocity and its uncertainty in $\kms$~from K22. Column (8) lists the measured background-subtracted specific intensity of the bow shock peak at 70~$\um$ in $\MJypersr$. It is worth noting that the \textit{HSO} PACS photometer has two imaging modes resulting in maps with 3\farcs2 or 1\farcs6 pixels. Converting fluxes to standard intensity units requires proper recognition of the imaging mode used. Most of the 70~$\um$ images were acquired in the parallel mode (\textit{i.e.}, a pixel size of 3\farcs2) but some notable targets\footnote{These targets are BS013$_\star$, BS344$_\star$ and BS386$_\star$.} were acquired with the 1\farcs6 pixel mode.\footnote{The bow shock studies of K18 and K19  erroneously used the 1\farcs6 pixel size for all images. Due to this difference, our measurements of $I_\nu$ are a factor of four lower than those presented in K18/K19 in most cases. Additional factors such as more-rigorous background subtraction may further reduce the tabulated $I_\nu$ compared to K18/K19.} Column (9) lists the logarithm of the irradiation parameter $U$, calculated from Equations~\ref{eq:three} and \ref{eq:four} (with $\gamma_\star=4.58$ from D21). Column (10) lists the terminal wind speed in $\kms$. Column (11) lists the measured chord length $\ell$ in arcseconds. Column (12) lists the $\log$ of the interpolated dust emissivity in 70~$\um$ for dust grains at the apex of the nebula $j_\nu$ in $\Jycmsrnuc$ from the D21 BPASS models at 10 Myr and solar metallicity. Column (13) lists the calculated ambient interstellar density computed from $n_a=I_\nu/(4j_\nu)$\footnote{This assumes the pre/post-shock density contrast ratio of four, appropriate to strong shocks \citep{LandauLifshitz1987}.} as in K18. Due to the difference in pixel sizes and background subtraction, our calculated densities are lower than those found in K18, by at least a factor of 4. Furthermore, K18 claimed 70~$\um$ detections in some cases where we determine only upper limit. We find $n_a$ ranges from 0.4--392~cm$^{-3}$ and has a median value of 2.2~cm$^{-3}$. This is markedly lower than the median $n_a$ of 16~cm$^{-3}$ reported by K18. The lower values of $n_a$ in this study challenge one of the findings of K18, in which the authors concluded that ambient densities of at least $n_a\gtrsim5~\text{cm}^{-3}$ were required to form bow shocks.

\citet{Henney2019a} provided a taxonomy of arc-like, infrared nebulae based on the coupling between gas and dust and the optical depth of the shell. They classified nebulae in which the gas and dust are coupled with low optical depths as wind-supported bow shocks, and they found that this scenario arises around stars with a broad range of stellar parameters and interstellar ambient densities $n_a<100~\text{cm}^{-3}$. \citet{Henney2019c} selected 23 sources from \citet{Kobulnicky17} and found all but one met the criteria of wind-supported bow shocks, confirming most arc-like nebulae are indeed bow shocks. Since we also select a subset of stars from K16, we expect most of our sources are wind-supported bow shocks as well. We note that four sources in Table~\ref{tab:BSstars} have $n_a>100~\text{cm}^{-3}$ (\textit{i.e.}, BS007, BS051, BS065, and BS067), so we consider these may not be wind-supported bow shocks. Hence, any measurements of $\mdot$ for these sources may not be reliable.\footnote{These may be bow waves, dust waves, or radiation bow shocks, in the nomenclature \citet{Henney2019a}.}

Some of the bow shock candidates cataloged in K16 may be interstellar bubbles, misidentified as bow shocks due to viewing angle effects \citep[\textit{e.g.}][]{Churchwell2006}. Additionally, episodic mass-loss events for evolved stars can produce nebulae that are not steady-state shock fronts formed from stellar winds. In light of this, it is unavoidable that some fraction of the nebulae cataloged in K16 and J19 are not formed from the stellar winds of a hot star. We aim to distinguish stellar bow shocks from morphologically similar phenomena on the basis of three criteria. First, we verify that our computed ambient densities (Column (13) of Table~\ref{tab:BSstars}) are consistent with wind-supported bow shock nebulae, and not other classes if arc-like infrared nebulae as discussed in \citet{Henney2019a}, by flagging as suspect objects with $n_a>100~\text{ cm}^{-3}$. Second, we ensure the morphology of the bow shock---from \textit{SST} 24~$\um$/\textit{WISE} 22~$\um$ images---is consistent with a steady-state stellar bow shock and discount ring-like and non-arc-like structures. We tabulate our results of morphology re-examination in Column (14) of Table~\ref{tab:BSstars}, with ``Y'' signifying arcuate structure and ``N'' signifying non-arcuate structure. Third, we search for indications of bubbles by searching for PAH emission along the edge of the nebula---consistent with the photo-dissociation front of a bubble---by examining \textit{SST} Infrared Array Camera 8.0~$\um$\citep{IRAC}/\textit{WISE} \texttt{W3} 12~$\um$ images. We note, however, that these passbands also contain warm dust emission. We tabulate our results of PAH indicators in Column (15) of Table~\ref{tab:BSstars}, with ``Y'' signifying possible PAH emission and ``N'' signifying no PAH emission.  We consider all three of these criteria for all objects, and assign a confidence indicator, either high (H) or low (L), based on these analyses. We tabulate the confidence indicator in Column (16) of Table~\ref{tab:BSstars}.

\begin{figure*}[htb!]
        \plottwo{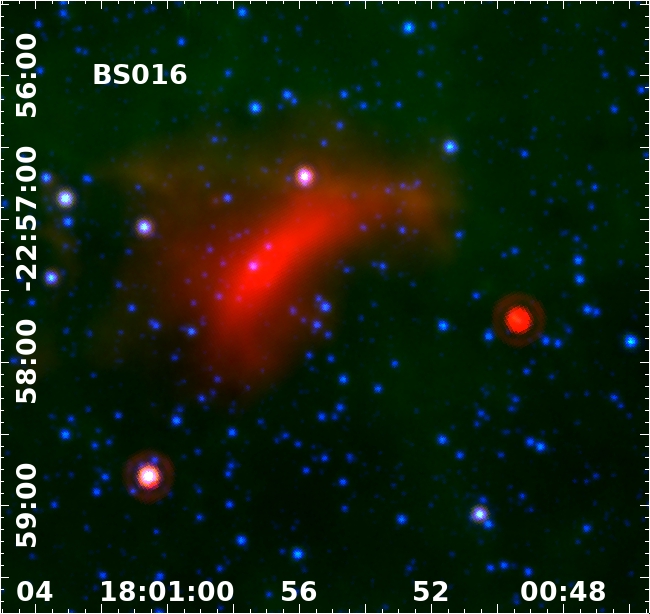}{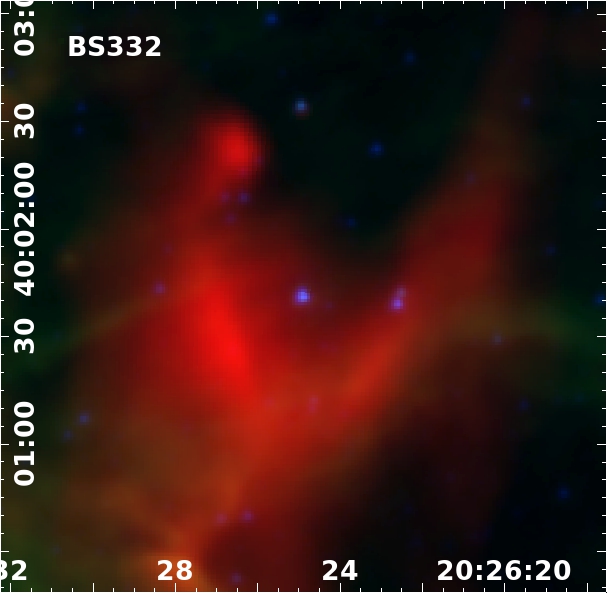}
        \plottwo{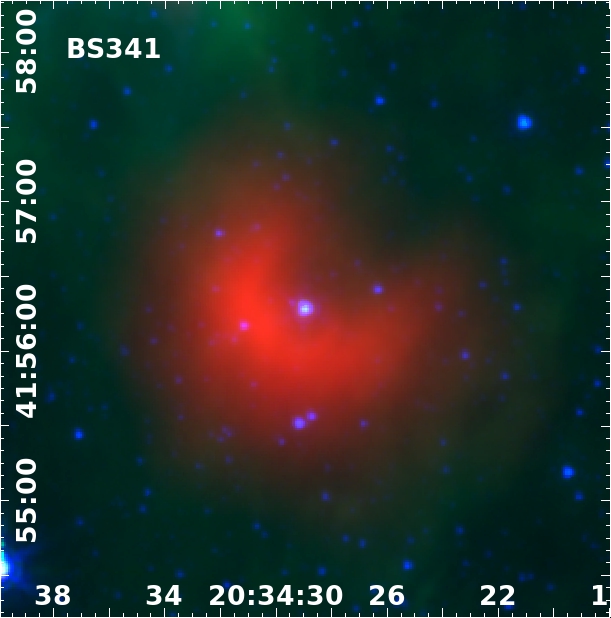}{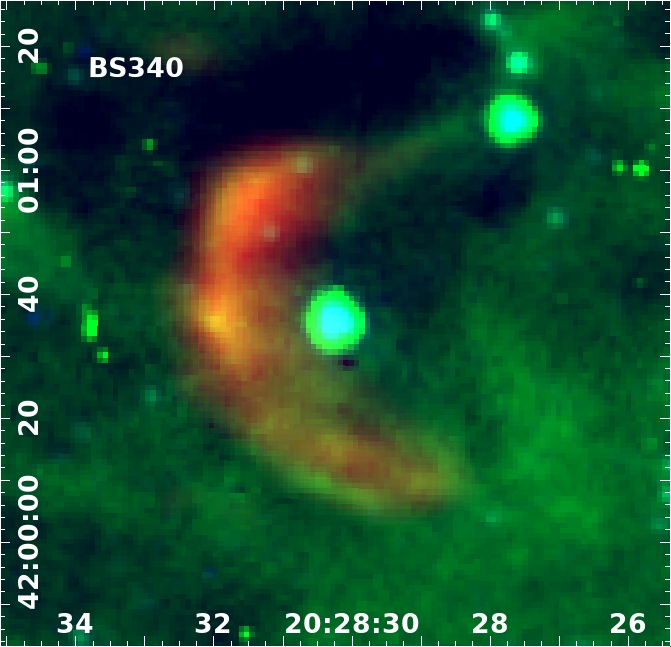}
        \caption{Four images of bow shock nebulae images from \textit{SST} with red/green/blue representing 24/8.0/4.5~$\um$ data. BS016 is a high-confidence nebula because of its morphology, BS332 is low confidence because of irregular morphology, 341 is high confidence because of its morphology and no apparent PAH emission along the shock, and 340 is low confidence because of the strong PAH emission along the bow shock arc.}
        \label{fig:outliers}
\end{figure*}



Figure~\ref{fig:outliers} displays infrared images of four selected bow shock nebulae that demonstrate the conditions for the morphology and PAH criteria. All panels in Figure~\ref{fig:outliers} display images from \textit{SST} with red/green/blue representing 24/8.0/4.5~$\um$ data. The top row presents a bow shock nebula that satisfies our morphology re-examination, and one that fails. BS016 (upper left) is well-defined in 24~$\um$, and has a clear, arcuate structure. There is no contaminating background emission at 24~$\um$, so we assign a ``Y'' for this bow shock's morphology indicator. BS332 (upper right), on the other hand, has a more dubious structure. There is a poorly defined bow shock arc surrounding the star, and there is also significant foreground/background signal in the 24~$\um$ data. The dubious morphology raises the possibility that this nebula does not represent a stellar-wind bow shock, so we assign an ``N'' flag for morphology. The bottom row of Figure~\ref{fig:outliers} presents a bow shock where there is no PAH emission (BS341) and one where there is possible PAH emission (BS340), indicated by the 8.0~$\um$ data (green). BS341 (bottom left) shows a clear bow shock structure in 24~$\um$ and no PAH signal coincident with the arc, so we assign an ``N'' flag for the PAH indicator for this nebula. Conversely, BS340 (bottom right) shows a clear bow shock structure and 8.0~$\um$ emission along the arc, giving it an orange/red appearance, so we assign a ``Y'' flag for its PAH indicator.

\begin{figure}[htb!]
        \plotone{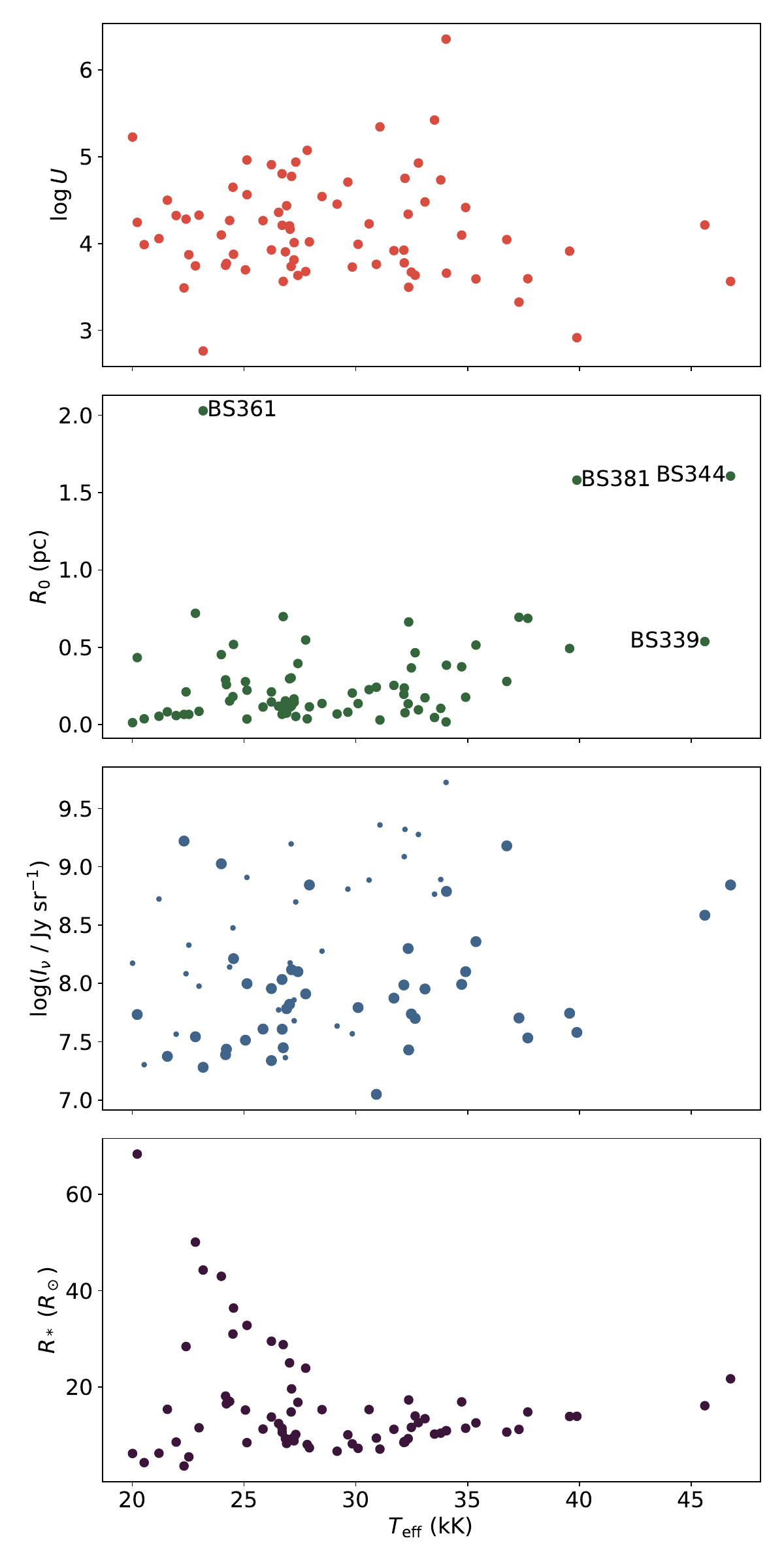}
        \caption{Distributions of the irradiation  energy parameter $\log U$, dust emissivity at 70~$\um$ $\log j_\nu$, specific intensity at 70~$\um$ $\log I_\nu$, and stellar radius in solar radii, versus stellar effective temperature $\teff$ in kK.}
        \label{fig:stacked_stellar}
\end{figure}




Figure~\ref{fig:stacked_stellar} displays a four panel plot of derived parameters versus stellar effective temperature $\teff$. Temperatures ranges from 19.8~kK on the cool end to 46.8~kK for the hottest star with median temperature 27.3~kK. The top panel plots the logarithm of the dimensionless irradiation parameter $U$. $\log U$ ranges from $2.8$ to $6.4$ with a median of $3.8$. $\log U$ does not correlate with stellar effective temperature, contrary to what might be expected given the greater radiant luminosity of hot stars. The lack of correlation in the top panel can be explained by the correlation in the second panel, which plots the standoff distance $R_0$ in parsecs versus $\teff$. $R_0$ in parsecs ranges from 0.01--2.03~pc with median 0.18~pc. The second panel shows that hotter stars have larger standoff distances, which will reduce the irradiation parameter $U$ at the apex of the bow shock. The larger standoff distances for hotter stars can be understood as a greater momentum flux of the wind. We note that there are several outliers in the second panel and label them. The third panel from the top plots the logarithm of the specific intensity at 70~$\um$ $I_\nu$ in $\Jysr$. Smaller points denote bowhsock nebulae that were not detectable above the background at 70~$\um$, which we treat as upper limits on $I_\nu$. We find no upper limit cases for stars hotter than 35~kK. $I_\nu$ ranges from $10^{7.0}$--$10^{9.7}$~$\Jysr$ with a median value of $10^{8.0}~\Jysr$. Specific intensity shows no correlation with effective temperature, unsurprisingly, for the same reason as the lack of correlation in the top panel. The bottom panel plots stellar radius versus temperature. The stellar radii range from 3.6--68.4~$\rsun$ with a median value of 11.5~$\rsun$. The data show a correlation between the two parameters, with a bifurcation below $\teff\approx27~\text{kK}$. The upper branch corresponds to evolved stars while the lower branch corresponds to main-sequence stars. At hotter temperatures the trends converge, signaling that the hottest stars are main-sequence stars.


\begin{figure}[htb!]
        \plotone{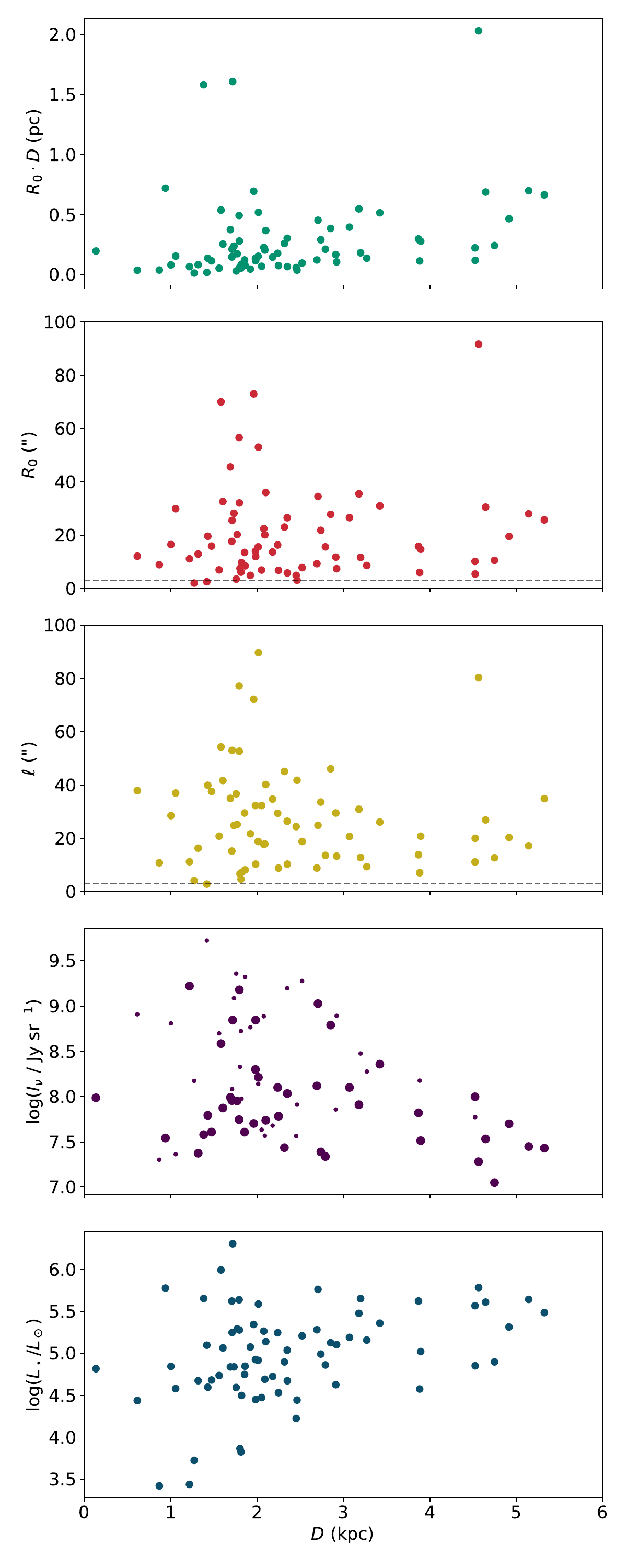}
        \caption{Distributions of the projected standoff distance $R_0$ in both parsecs and arseconds, chord length $\ell$ in arcseconds, log surface brightness $I_\nu$ in $\Jysr$, and log stellar luminosity $\lstar$ in solar luminosity units versus the parallactic distance $D$, in kpc.}
        \label{fig:stacked_morphology}
\end{figure}




Figure~\ref{fig:stacked_morphology} is a five-panel plot of derived parameters versus parallactic distance $D$ to consider whether there are any biases with distance. Excluding one object at 11.5~kpc, distances range from 130~pc to 5.3~kpc with a median distance of 2.1~kpc. The top panel plots the projected standoff distance $\rnaught$ in parsecs, which ranges from 0.01--2.03~pc, and has a median of 0.18~pc. The second panel plots the projected standoff distance $\rnaught$ in arcseconds versus distance. Excluding three anomalously large bow shock nebulae with projected standoff distances exceeding 100$"$, $\rnaught$ ranges from 2$\farcs$0--91$\farcs$7 with a median of 15$\farcs$8. Most bow shocks were identified in K16 at 24~$\um$ from \textit{SST} images, which have a nominal resolution of $6\farcs0$, and we expect bow shock nebulae with standoff distances $R_0\lesssim3\arcsec$ to be systematically underrepresented in the K16 master sample. We plot this minimum resolvable projected standoff distance as a horizontal dashed black line in the second panel. Consequently, the median bow shock nebula with an $\rnaught=0.18~\text{pc}$ should be resolvable out to 12~kpc, but the smallest ones ($\rnaught=0.01~\text{pc}$) would become missing beyond 1.2~kpc. The third panel plots the chord length of the bow shock nebulae. Excluding large bow shock nebulae with $\ell >100\arcsec$, we find chord lengths ranging from 2$\farcs$8--90$\arcsec$ with median of $24\farcs9$. In a similar fashion as the second panel, we plot the minimum measurable chord length due to the resolution of \textit{SST} at 24~$\um$ as a horizontal dashed black line. The fourth panel plots the logarithm of the specific intensity at 70~$\um$ in Jy sr$^{-1}$. Smaller points represent nondetections (which we treat as upper limits on $I_\nu$) and larger points denote detections above the background. $\log I_\nu$ ranges from 7.1--9.7 with a median of 8.0. At distances $D<3.5~\text{ kpc}$ the full range of specific intensities is present. Nearly 2/3 of these nearby bow shocks with $\log I_\nu>8.5$ are upper limits. However, beyond 3.5~kpc, there is a paucity of sources with $\log I_\nu>8.5$. We cannot attribute this lack of high $I_\nu$, distant sources to confusing foreground and background contamination, since the most-distant sources are mostly confirmed detections. We also cannot attribute this lack to beam dilution as these nebulae exceed the angular resolution of both \textit{SST} (6$\arcsec$) and \textit{HSO} (5\farcs5)---by factors of several---having an angular size of 20$\arcsec$, consistent with the rest of the sample. The absence of high $I_\nu$ distant targets in our sample is apparent, although do not have an explanation. The bottom panel plots the logarithm of stellar luminosity $\lstar$ in solar luminosity units. $\log\lstar$ ranges from 3.4--6.3, with median 5.0. The lack of low-luminosity stars beyond 2~kpc can readily be explained by the combination of distance and extinction, which act to push these stars beyond our spectroscopy limit of Gaia $B_P\lesssim15$ mag. An extended bow shock study would need more-sensitive blue-optical spectra to include these fainter bow shock-driving stars.

\section{$\dot{M}$ for bow shock stars}
\label{sec:results}

\startlongtable
\begin{deluxetable*}{lcccccccc}
	\tablecaption{$\mdot$ for bow shock stars}\label{tab:mdots}
	\tabletypesize{\scriptsize}
	\tablehead{
		ID & $\mstar$ & $\log\left(\lstar/L_\odot\right)$ & $V_{2D}$ & $\log\mdot$ & $\log\mdot_{20}$ & $\sigma_{\log\mdot}$ & $\log\mdot_{\text{Vink}}$ & $\log\mdot_{\text{Bjorklund}}$ \\
		& ($M_\odot$) & (dex) & $\left(\kms\right)$ & $\left(\moperyear\right)$ & $\left(\moperyear\right)$ & $\left(\moperyear\right)$ & $\left(\moperyear\right)$ & $\left(\moperyear\right)$ \\
		(1) & (2) & (3) & (4) & (5) & (6) & (7) & (8) &(9)
	}
	\startdata
		BS001$_{\star}$ & 53.1 & 5.13 & 19 & -6.90 & -6.85 & 0.19 & -6.98 & -7.74\\
		BS007$_{\star}$ & 7.8 & 3.44 & 11 & -7.05 & -6.53 & 0.17 & -9.15 & -10.94\\
		BS010$_{\star}$ & 13.6 & 4.44 & 19 & -10.18 & -10.14 & 0.15 & -8.01 & -8.69\\
		BS013$_{\star}$ & 14.9 & 4.82 & 23 & -7.77 & -7.77 & 0.17 & -7.01 & -7.72\\
		BS016$_{\star}$ & 34.8 & 5.19 & 5 & -8.23 & -7.02 & 0.36 & -6.93 & -7.52\\
		BS019$_{\star}$ & 78.0 & 5.65 & 26 & -7.29 & -7.29 & 0.13 & -5.54 & -7.03\\
		BS026$_{\star}$ & 72.0 & 5.16 & 8 & -8.71 & -7.92 & 0.25 & -7.34 & -8.03\\
		BS027$_{\star}$ & 9.4 & 3.86 & 6 & -8.62 & -7.58 & 0.32 & -8.31 & -10.03\\
		BS043$_{\star}$ & 14.2 & 4.53 & 6 & -9.71 & -8.66 & 0.32 & -6.99 & -8.54\\
		BS051$_{\star}$ & 95.0 & 5.10 & 3 & -9.83 & -8.18 & 0.31 & -7.38 & -8.19\\
		BS056$_{\star}$ & 146.0 & 5.64 & 32 & -6.86 & -6.86 & 0.88 & -5.89 & -7.38\\
		BS061$_{\star}$ & 64.9 & 5.26 & 10 & -7.47 & -6.86 & 0.21 & -6.93 & -7.64\\
		BS063$_{\star}$ & 58.0 & 4.59 & 5 & -10.07 & -8.86 & 0.21 & -8.32 & -9.16\\
		BS065$_{\star}$ & 9.1 & 3.82 & 3 & -8.88 & -7.23 & 0.44 & -8.40 & -10.15\\
		BS067$_{\star}$ & 59.6 & 4.85 & 4 & -8.65 & -7.25 & 0.26 & -7.73 & -8.54\\
		BS150$_{\star}$ & 16.1 & 4.44 & 8 & -9.19 & -8.39 & 0.27 & -7.30 & -8.90\\
		BS200$_{\star}$ & 80.0 & 5.57 & 34 & -7.43 & -7.43 & 0.36 & -5.73 & -7.23\\
		BS255$_{\star}$ & 8.5 & 3.72 & 19 & -9.32 & -9.28 & 0.15 & -8.61 & -10.41\\
		BS288$_{\star}$ & 18.5 & 4.63 & 11 & -8.76 & -8.24 & 0.20 & -7.82 & -8.47\\
		BS300$_{\star}$ & 17.0 & 4.47 & 4 & -10.87 & -9.47 & 0.25 & -7.98 & -8.71\\
		BS302$_{\star}$ & 12.9 & 4.22 & 7 & -10.24 & -9.33 & 0.28 & -7.71 & -9.40\\
		BS303$_{\star}$ & 85.0 & 5.90 & 52 & -6.62 & -6.62 & 0.88 & -5.14 & -6.68\\
		BS305$_{\star}$ & 93.0 & 5.08 & 9 & -9.67 & -8.98 & 0.50 & -7.43 & -8.24\\
		BS306$_{\star}$ & 19.1 & 4.72 & 7 & -9.49 & -8.58 & 0.26 & -7.62 & -8.25\\
		BS314$_{\star}$ & 31.0 & 4.85 & 11 & -9.40 & -8.88 & 0.20 & -6.74 & -8.28\\
		BS320$_{\star}$ & 54.5 & 5.48 & 7 & -8.07 & -7.15 & 0.28 & -6.53 & -7.11\\
		BS321$_{\star}$ & 65.1 & 4.90 & 1 & -11.24 & -8.64 & 0.88 & -7.72 & -8.51\\
		BS322$_{\star}$ & 74.0 & 5.31 & 3 & -9.26 & -7.61 & 0.32 & -6.81 & -7.55\\
		BS324$_{\star}$ & 114.0 & 5.49 & 30 & -7.41 & -7.41 & 0.14 & -6.69 & -7.42\\
		BS325$_{\star}$ & 18.7 & 4.67 & 12 & -9.28 & -8.83 & 0.19 & -6.94 & -8.57\\
		BS330$_{\star}$ & 30.9 & 5.25 & 9 & -8.57 & -7.87 & 0.16 & -5.94 & -7.49\\
		BS331$_{\star}$ & 89.0 & 4.84 & 4 & -8.95 & -7.56 & 0.45 & -7.89 & -8.75\\
		BS332$_{\star}$ & 15.8 & 4.84 & 5 & -7.71 & -6.51 & 0.21 & -6.99 & -7.70\\
		BS335$_{\star}$ & 21.0 & 4.92 & 12 & -8.24 & -7.80 & 0.19 & -6.41 & -7.96\\
		BS339$_{\star}$ & 89.0 & 6.00 & 77 & -5.87 & -5.87 & 0.13 & -5.23 & -5.73\\
		BS340$_{\star}$ & 17.4 & 4.68 & 19 & -8.84 & -8.80 & 0.15 & -6.79 & -8.34\\
		BS341$_{\star}$ & 91.0 & 5.29 & 11 & -8.80 & -8.28 & 0.20 & -6.96 & -7.72\\
		BS344$_{\star}$ & 104.0 & 6.31 & 45 & -5.40 & -5.40 & 0.13 & -4.64 & -5.06\\
		BS350$_{\star}$ & 22.1 & 4.90 & 10 & -8.84 & -8.24 & 0.23 & -6.48 & -8.04\\
		BS353$_{\star}$ & 73.0 & 5.61 & 67 & -6.56 & -6.56 & 0.14 & -6.02 & -6.70\\
		BS356$_{\star}$ & 19.3 & 4.86 & 10 & -8.82 & -8.22 & 0.21 & -6.45 & -7.96\\
		BS357$_{\star}$ & 23.8 & 4.99 & 8 & -8.91 & -8.12 & 0.25 & -6.32 & -7.86\\
		BS358$_{\star}$ & 6.8 & 3.42 & 2 & -10.94 & -8.94 & 0.45 & -9.15 & -10.97\\
		BS361$_{\star}$ & 48.6 & 5.78 & 29 & -6.18 & -6.18 & 0.14 & -5.00 & -6.47\\
		BS362$_{\star}$ & 59.4 & 5.02 & 27 & -7.72 & -7.72 & 0.14 & -6.77 & -8.35\\
		BS365$_{\star}$ & 16.0 & 4.58 & 11 & -9.01 & -8.49 & 0.20 & -6.95 & -8.50\\
		BS367$_{\star}$ & 58.9 & 5.14 & 8 & -8.52 & -7.72 & 0.25 & -7.07 & -7.82\\
		BS368$_{\star}$ & 53.4 & 4.93 & 8 & -9.03 & -8.23 & 0.16 & -7.49 & -8.27\\
		BS369$_{\star}$ & 68.0 & 5.34 & 9 & -8.02 & -7.33 & 0.32 & -6.57 & -7.30\\
		BS381$_{\star}$ & 69.0 & 5.65 & 11 & -7.18 & -6.67 & 0.27 & -5.86 & -6.50\\
		BS384$_{\star}$ & 81.0 & 5.36 & 4 & -8.25 & -6.85 & 0.45 & -6.67 & -7.42\\
		BS386$_{\star}$ & 39.5 & 5.59 & 48 & -6.40 & -6.40 & 0.13 & -5.29 & -6.75\\
		BS407$_{\star}$ & 25.8 & 5.21 & 10 & -8.28 & -7.68 & 0.21 & -6.43 & -7.11\\
		BS409$_{\star}$ & 50.9 & 5.10 & 19 & -7.88 & -7.84 & 0.16 & -7.02 & -7.78\\
		BS414$_{\star}$ & 15.7 & 4.67 & 7 & -9.65 & -8.74 & 0.18 & -6.74 & -8.27\\
		BS634$_{\star}$ & 65.9 & 5.06 & 9 & -8.61 & -7.91 & 0.22 & -7.33 & -8.09\\
		BS640$_{\star}$ & 71.0 & 4.74 & 12 & -8.92 & -8.48 & 0.19 & -8.34 & -9.07\\
		BS667$_{\star}$ & 55.8 & 4.85 & 29 & -7.60 & -7.60 & 0.13 & -7.82 & -8.57\\
		BS673$_{\star}$ & 14.4 & 4.57 & 21 & -7.88 & -7.88 & 0.18 & -7.81 & -8.43\\
		BS687$_{\star}$ & 78.0 & 5.28 & 20 & -8.13 & -8.13 & 0.17 & -7.21 & -7.83\\
		BS692$_{\star}$ & 98.0 & 5.28 & 17 & -6.98 & -6.84 & 0.12 & -6.93 & -7.70\\
		BS700$_{\star}$ & 84.0 & 5.62 & 16 & -7.94 & -7.75 & 0.21 & -6.51 & -7.06\\
		BS705$_{\star}$ & 25.5 & 5.04 & 24 & -6.04 & -6.04 & 0.14 & -7.11 & -7.69\\
		2G0063185-0066622$_{\star}$ & 68.0 & 5.76 & 8 & -6.92 & -6.12 & 0.25 & -5.23 & -6.70\\
		2G0065094-0061344$_{\star}$ & 15.1 & 4.45 & 5 & -8.27 & -7.07 & 0.54 & -8.05 & -8.74\\
		2G0118310-0062990$_{\star}$ & 15.6 & 4.50 & 5 & -9.17 & -7.97 & 0.37 & -7.19 & -8.82\\
		2G0592097+0012519$_{\star}$ & 145.0 & 5.62 & 4 & -9.59 & -8.19 & 0.48 & -5.94 & -7.44\\
		2G0600615-0018340$_{\star}$ & 20.3 & 4.69 & 8 & -8.81 & -8.02 & 0.16 & -7.57 & -8.29\\
		2G0770694+0193568$_{\star}$ & 19.2 & 4.60 & 4 & -9.86 & -8.47 & 0.45 & -7.73 & -8.47\\
		2G1045666+0128085$_{\star}$ & 111.0 & 5.25 & 4 & -9.69 & -8.29 & 0.45 & -7.12 & -7.91\\
		$\kappa$ Cas & 60.0 & 5.78 & 23 & -7.06 & -7.06 & 0.23 & -5.15 & -6.63\\
		2G3525610+0001887$_{\star}$ & 21.2 & 4.75 & 16 & -8.93 & -8.74 & 0.21 & -6.74 & -8.28\\
		2G3533026+0008095$_{\star}$ & 67.0 & 5.64 & 24 & -7.77 & -7.77 & 0.12 & -5.88 & -6.53\\
	\enddata
	\tablecomments{Column (1): K16 or J19 identifier. Column (2):  Stellar mass in solar masses. Column (3): $\log$ of stellar luminosity in solar luminosity units. Column (4): 2D peculiar space velocity in $\kms$. Column (5): $\log$ of the computed mass-loss rate in $\moperyear$. Columns (6): $\log$ of the corrected mass-loss rate in $\moperyear$. Column (7) The random uncertainty in the computed mass-loss rates in columns (5) and (6) from error propagation. Column (8): The predicted mass-loss rate from the \citet{Vink2001} relation. Column (9): The predicted mass-loss rate from the \citet{Bjorklund2023} relation.}
\end{deluxetable*}

Table~\ref{tab:mdots} presents the results of our computations of mass-loss rates for \numstars~bow shock stars. Column (1) lists the bow shock identifier, either from the K16 or J19 catalogs. Column (2) lists the stellar mass of the bow shock central star in solar masses. Column (3) lists the logarithm of stellar luminosity of the bow shock central star in solar luminosity units. Column (4) presents the 2D peculiar space velocity of the bow shock central star in $\kms$ from K22 and calculated from proper-motion data. Column (5) presents the logarithm of the calculated mass-loss rate of the bow shock central star using the parameters listed in Table~\ref{tab:BSstars} and Equation~\ref{eq:two}. Column (6) presents the corrected mass-loss rate $\mdot_{20}$---assuming $V_a$ takes a minimum value of 20~$\kms$---which is expounded upon later in this section. Column (7) presents the random uncertainty in the calculated mass-loss rates in columns (5) and (6) using standard error propagation and given uncertainties of parameters in Equation~\ref{eq:two}. Columns (8) and (9) list the predicted mass-loss rate of each tabulated star using the analytical formulae of \citet[Equations 24 and 25][V01 hereafter]{Vink2001} and \citet[Equation 7][B23 hereafter]{Bjorklund2023}.

\begin{figure*}[htb!]
        \plotone{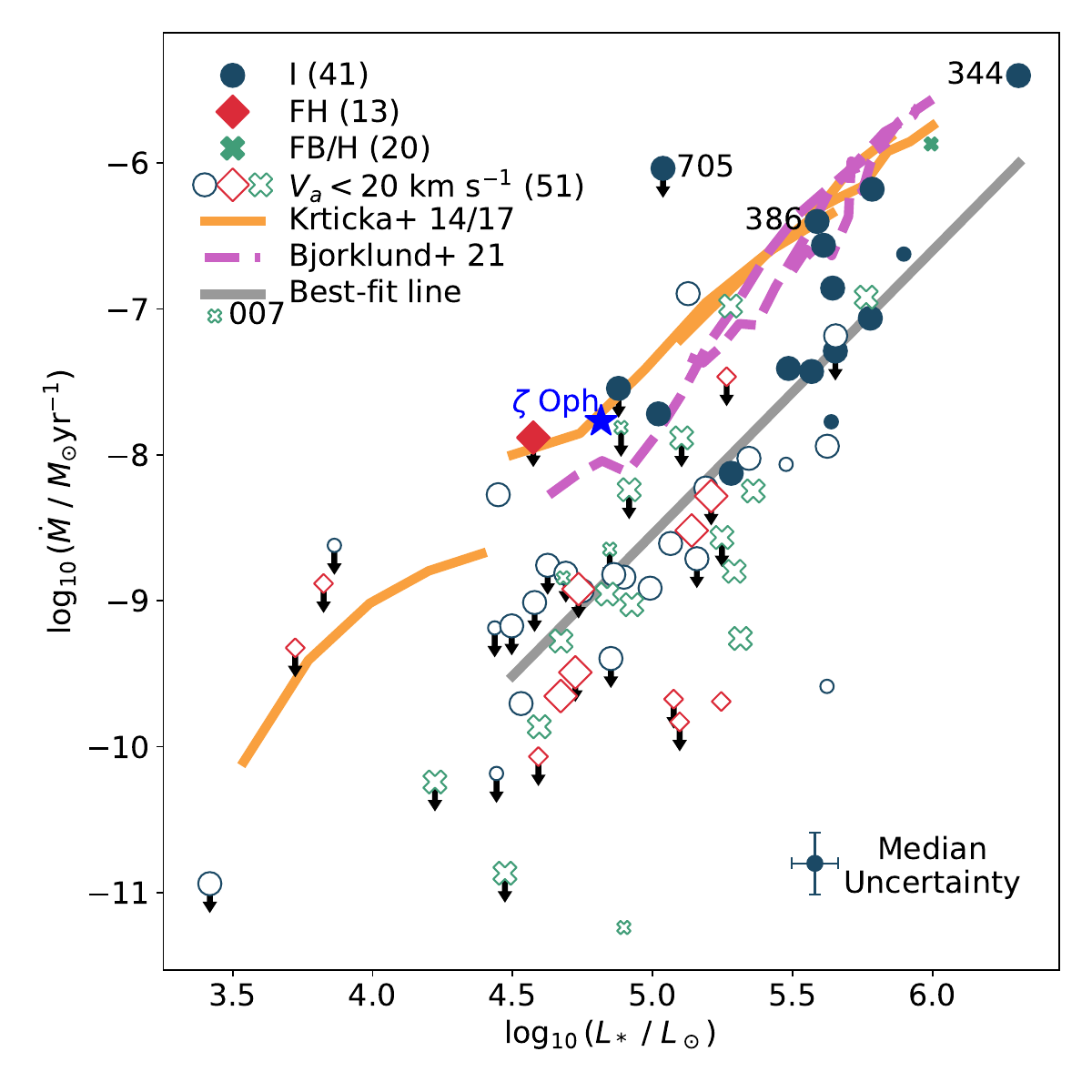}
        \caption{Mass-loss rates of bow shock stars plotted versus stellar luminosity. Symbols denote local environment as in the legend. Hollow points denote slow-moving stars. Large points denote high-confidence measurements. Orange lines signify prescriptions from \citet{Krticka2014} and \citet{KrtickaandKubalt2017}. The magenta track marks the models from \citet{Bjorklund2023}. The black line marks the best-fit trend for measurements of $\mdot$ for stars with $\log\lstar>4.5$.}
        \label{fig:mass_loss_full}
\end{figure*}










Figure~\ref{fig:mass_loss_full} plots the logarithm of the mass-loss rates, in $\moperyear$, versus the logarithm of stellar luminosity, in solar luminosity units, for \numstars~bow shock stars. Following the nomenclature of K16, we classify targets according to their apparent local environment. Blue circles represent isolated bow shock nebula (I, 41), red diamonds denote bow shock nebulae that face \ion{H}{2} regions (FH, 13), and green $\times$'s represent bow shock nebulae that either face bright-rimmed clouds or lie within \ion{H}{2} regions (FB/H, 20). These latter two flags are grouped together given the ambiguity in classification for a bow shock nebula on the edge of an \ion{H}{2} region which faces a photodissociation front. K22 found that for runaway bow shock stars ($V_{2D}\geq25~\kms$) the morphological position angle of the IR nebula is highly aligned with the star's peculiar motion vector, while for non-runaway stars ($V_{2D}<25~\kms$) the difference between these two vectors ($\Delta\text{PA}$ in the notation of K22) is isotropic. The authors concluded that in the non-runaway cases---the majority of bow shock stars (76\%)---bulk ISM flows played a more prominent role in forming the bow shock than the star's space velocity. This can lead to an under-measurement of $\mdot$ since the bow shock technique assumes the star's peculiar 2D space velocity is equivalent to the star's velocity relative to the ambient material ($V_a$ in Equation~\ref{eq:two}). This is corroborated by \citet{vandenEijndelandKaper2026}, who found that ignoring the effect of bulk ISM motions for stars with low space velocities (less than about 15 $\kms$) leads to a systematic underestimate in $\dot{M}$. Based on these findings, we adopt hollow symbols to mark bow shock stars with space velocities $V_{2D}<20~\kms$ (\numslow\ of \numstars\ targets). Downward arrows mark upper limits for cases where no 70~$\um$ emission is detected from the bow shock above the background (30 targets). The blue star signifies the measurement for the runaway star $\zeta$ Oph. Orange lines signify theoretical prescriptions for B-main-sequence and O-dwarf, giant, and supergiant stars \citep[][]{Krticka2014, KrtickaandKubalt2017}. Magenta lines signify the solar-metallicity model calculations of $\mdot$ for dwarf, giant, and supergiant stars from \citet[][tabulated in Table A.1 in their Appendix]{Bjorklund2021}. The semi-transparent black line marks the best-fit trend to the data for targets with $\log \lstar>4.5$. Text labels mark some outliers discussed in the text below. The point at the lower right illustrates the median random uncertainties in $\mdot$ and $\lstar$. 





The points in Figure~\ref{fig:mass_loss_full} show a strong correlation between mass-loss rate and luminosity. The slope of the best-fit trend has the same slope as the models but is offset downward by 0.8--1.0~dex. The upper envelope of the data---mostly consisting of large solid points indicating high-confidence measurements moving at substantial velocities---agrees well with both the \citet{Bjorklund2021} and the \citet{KrtickaandKubalt2017}/\citet{Krticka2014} predictions. The hollow points scatter below these models by 1--2~dex. Measurements of $\mdot$ for each of the three classes of environments are consistent with the distribution of the full sample, indicating that measurements of $\mdot$ are not significantly impacted by environment. Below $\log\lstar<5$, the sample is increasingly dominated by upper limits. The upper-limits follow the same correlation as the measurements with detections, although with a higher dispersion. Low-confidence objects (small points) exhibit the same trend as high-confidence nebulae (large symbols) albeit with a larger dispersion. Measurements for $\mdot$ have an RMS of 0.85~dex about the best-fit line.




The prototypical bow shock star $\zeta$ Ophiuchi (BS$013_\star$) lies in the middle of the plot (blue star). $\zeta$ Oph is situated along the top of the distribution, helping define the upper envelope. It shows excellent agreement with the \citet{KrtickaandKubalt2017} and \citet{Bjorklund2021} models. We calculate a mass-loss rate of $\dot{M}=1.7\times10^{-8}~\moperyear$ which is 40\% higher than the measurement of K19. Our measurement for $\dot{M}$ is within 10\% of the B23 analytical prescription\footnote{This amounts to a factor of 1.7 lower than Equation~2 in \citet[][$\dot{M}=2.88\times10^{-8}~\moperyear$.]{Krticka2025}}. The new measurement and B23 prediction are both lower than the V01 of $9.77\times10^{-8}~\moperyear$. Measurements of $\mdot$ for $\zeta$ Oph vary greatly in the literature, and our measurement lies in the middle of other values. On the low end, \citet{Marcolino2009} measured a mass-loss rate of $1.58\times10^{-9}~\moperyear$, which is a factor of 10 lower than our value. X-ray data yielded a similar value of $1.50\times10^{-9}~\moperyear$ \citep{Cohen2014}. Towards the higher end, the radio-continuum analysis of \citet{Rubio-Diez2022} yielded an upper limit of $\dot{M}<1.8\times10^{-7}~\moperyear$, a factor of 10 higher than our value.

BS344$_\star$ (BD+43 3654) is a well-studied runaway star ($V_{2D}=45~\kms$) and is the most-luminous star in our sample ($\log\lstar=6.3$, $\log\mdot=-5.4$). We measure a mass-loss rate of $\mdot=4.0\times10^{-6}~\moperyear$. This is a factor of five lower than K19 on account of the reduced 70~$\um$ specific intensity from rigorous background subtraction. BS344$_\star$ is more luminous than any of the plotted models but shows excellent agreement with an extrapolation of either the \citet{Bjorklund2021} or \citet{KrtickaandKubalt2017} predictions. Our measurement is a factor of seven lower than the V01 prediction and is a factor of two lower than the B23 prediction.

BS386$_\star$ (Vela X-1) is a runaway compact object+B1 supergiant binary system ($\log\lstar=5.6$, $\log\mdot=-7.0$). We measure a mass-loss rate of $\mdot=1.0\times10^{-7}~\moperyear$, which agrees well with both the plotted \citet{KrtickaandKubalt2017} and \citet{Bjorklund2021} predictions. Our value is a factor of 50 times lower than the V01, and a factor of 2 lower than the B23 analytical formulae. In their review of Vela X-1, \citet{Kretschmar2021} provided an overview of the different measurements of mass loss by stellar winds for the donor star of this system as well as the method used in these studies (see their Table~7). X-ray- and UV-based measurements yield a range of mass-loss rates, taking values from $2.7\times10^{-7}~\moperyear$ on the low end, up to $<5.3\times10^{-6}~\moperyear$ on the high end. Our measurement is consistent with these figures and lies on lower end of published values.

The measurement of $\mdot$ for BS$007_\star$ ($\log\lstar=3.4$, $\log\mdot=-7.1$) stands out as conspicuously high for its low luminosity. There is an apparent bright source of 70~$\um$ emission at the location of the 24~$\um$ peak. However, it has the lowest 24/70 flux ratio of any bow shock measured in Table~1 in \citet{Kobulnicky17}, indicating that there may be 70~$\um$ contamination along the same sight line. BS007$_\star$ lies within 1$^\circ$ of the Galactic plane and 3$^\circ$ from Galactic center. In such a crowded field, it is likely that there are overlapping foreground or background structures that contaminate the 70~$\um$ signal, inflating our measurement of $\dot{M}$. This high 70~$\um$ intensity leads to an anomalously high ambient density ($n_a=392~\text{cm}^{-3}$, the highest in our sample), suggesting that the nebula may not represent a stellar-wind bow shock, which casts doubt on our measurement of $\mdot$ using the bow shock method.

BS705$_\star$ ($\log\lstar=5.0$, $\log\mdot=-6.0$) also appears as an outlier on Figure~\ref{fig:mass_loss_full}. The bow shock nebula shows no detectable signal above the background at 70~$\um$, meaning the derived mass-loss rate should be regarded as an upper limit. The star lies within a degree of Galactic center and similar to BS$007_\star$, there is likely foreground and background contaminating emission at 70~$\um$. We measure a specific intensity at 70~$\um$ $I_\nu$ of $1.57\times10^9~\Jysr$, among the brightest in our sample. As a consequence, it has among the lowest $F_{24}/F_{70}$ flux ratios \citep{Kobulnicky17} and highest ambient densities of $n_a=60.1~\text{cm}^{-3}$---however, still within the expected range of stellar bow shock nebulae ($n_a<100~\text{ cm}^{-3}$). We attribute background/foreground contamination as the source of the high $\mdot$ but find no reason to discount this measurement otherwise. As an upper limit, this measurement is consistent with the theoretical predictions.




The majority of objects in the sample (\numslow\ of \numstars) have space velocities less than 20~$\kms$ (open symbols in Figure~\ref{fig:mass_loss_full}; we will call these the ``low-velocity stars''). These group towards the lower portion of the plot and lie an order of magnitude below the upper envelope of measurements. Compared to the models, the hollow symbols lie as much as 1--2 orders of magnitude below the theoretical predictions. Since ISM flows of unconstrained velocity are likely the mechanism for forming these bow shocks (as found in K22), we regard our calculated mass-loss rates---which rely upon space velocities computed from stellar proper-motion data---as lower limits. Low-velocity stars with 70~$\um$ upper limits seem to follow the same trend as the other low-velocity stars. Evidently, the impact of having 70~$\um$ non-detections is less important than the impact of having anomalously low $V_a$. This can be understood by reference to Equation~\ref{eq:two}. The former parameter enters only linearly while $V_a$ takes a squared dependence, making it the largest source of uncertainty (indeed, systematic uncertainty!) in computing $\mdot$.

\begin{figure}[htb!]
        \plotone{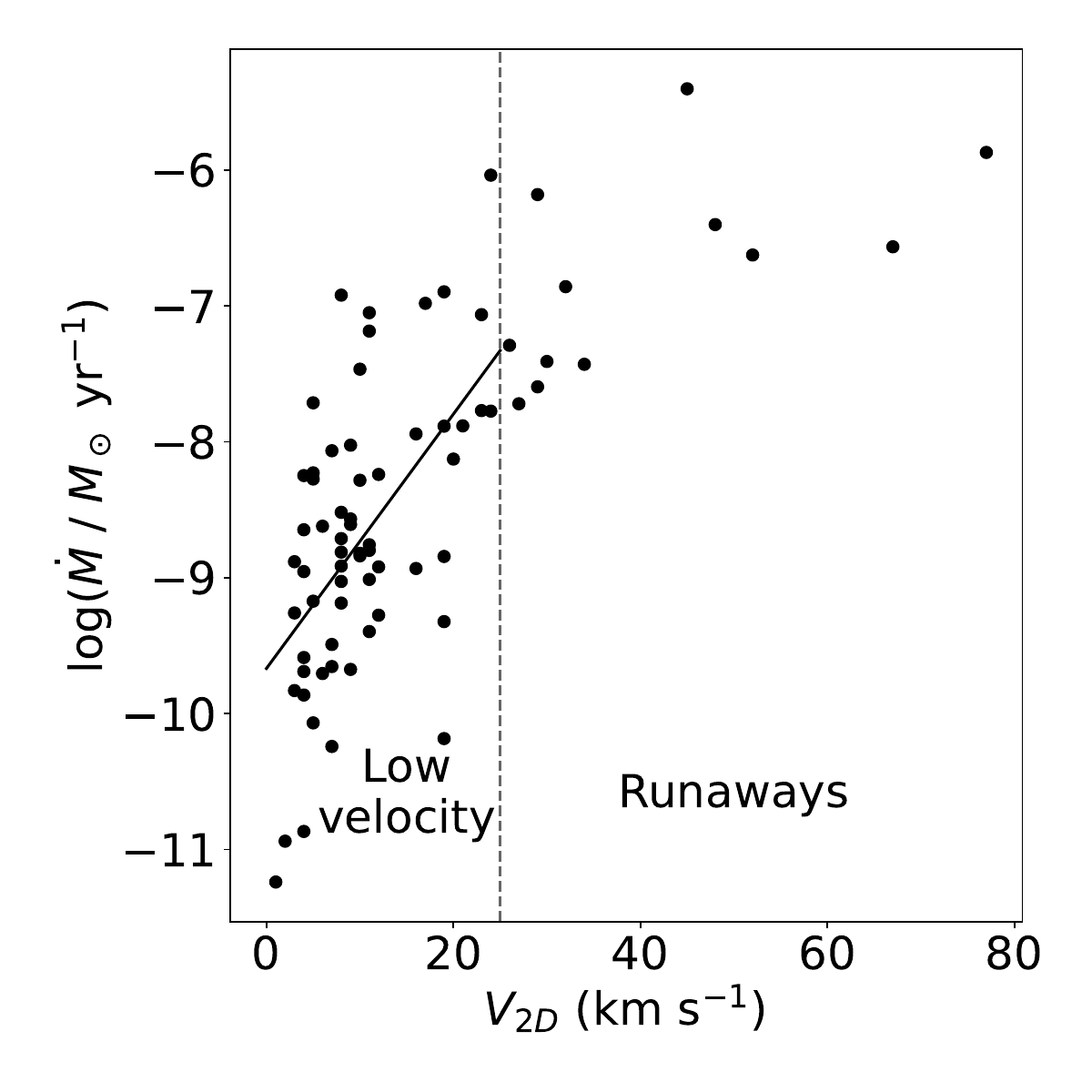}
        \caption{Logarithm of mass-loss rate, in solar masses per year, versus peculiar 2D space velocity, in $\kms$. The dashed vertical line shows the runaway velocity of 25~$\kms$ and the black line shows the best-fit linear relation for the low-velocity stars.}
        \label{fig:v_mdot_correlation}
\end{figure}



\citet{vandenEijndelandKaper2026} found that mass-loss rates calculated from low-velocity stars using the bow shock technique are systematically underestimated when ISM bulk flows are ignored. To investigate the potential reasons for the offset observed for low-velocity stars, we plot in Figure~\ref{fig:v_mdot_correlation} mass-loss rate versus the star's 2D peculiar space velocity in the same manner as Figure~10 in \citet{vandenEijndelandKaper2026}. The dashed vertical line denotes the threshold for runaway stars of 25~$\kms$ as used in K22, and the black line shows the best-fit trend for the low-velocity stars. The runaway stars on the right side of the panel show minimal-or-no correlation between mass-loss rate and $V_{2D}$ while the low-velocity stars show a positive correlation. This correlation between $\mdot$ and $V_{2D}$ on the left-hand side is unexpected and unphysical, unless perhaps lower-mass stars have lower space velocities. Said another way, this trend can be explained if low-mass OB stars are preferentially produced with low space-velocities. \citet{vandenEijndelandKaper2026} raise this as a possibility, and we investigate this question in the following figure.

\begin{figure}[htb!]
        \plotone{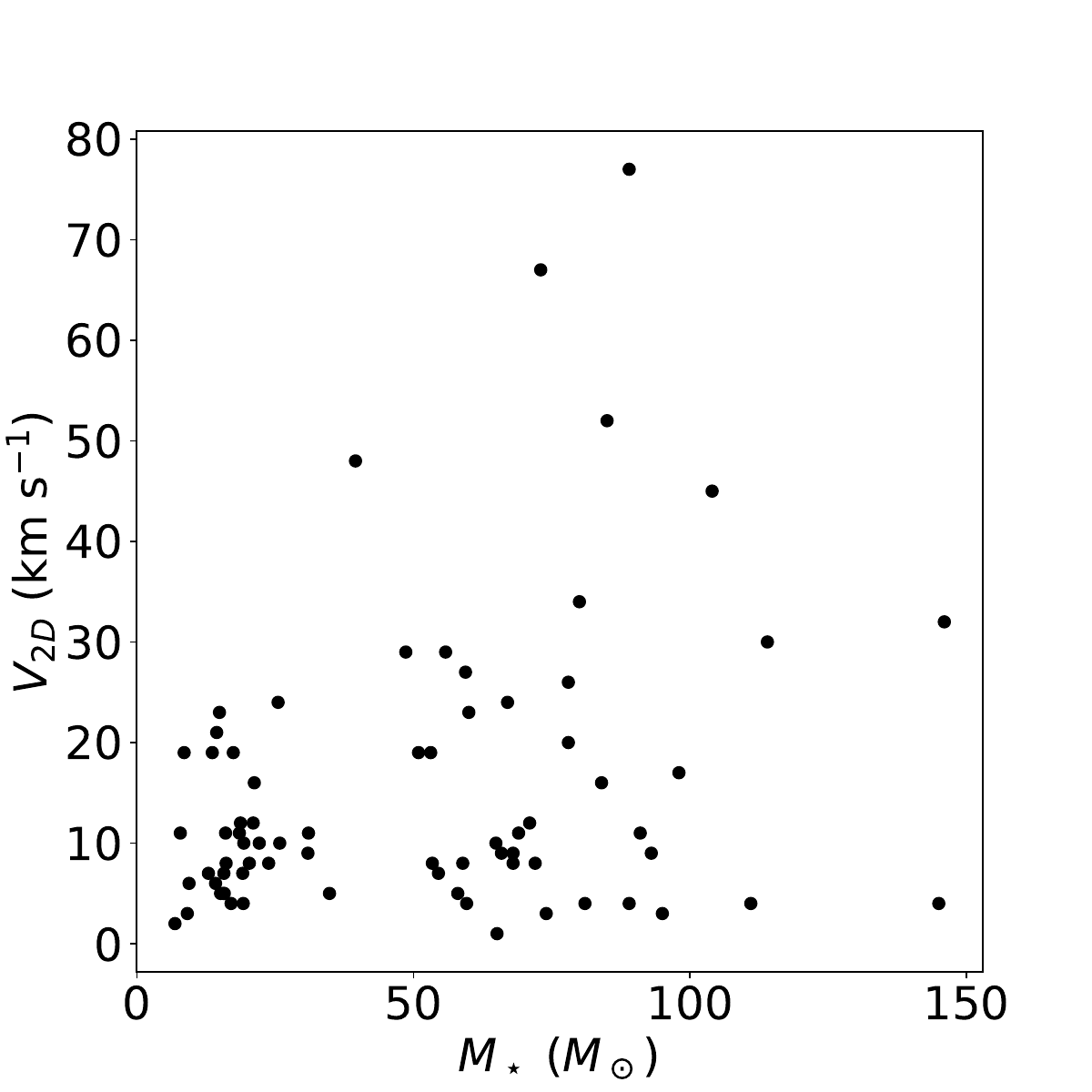}
        \caption{Peculiar 2D space velocity, in $\kms$, versus stellar mass, in solar masses.}
        \label{fig:mass_velocity_correlation}
\end{figure}

Figure~\ref{fig:mass_velocity_correlation} plots the peculiar 2D space velocity, in $\kms$, versus stellar mass, in solar masses. We find no correlation between space velocity and stellar mass. Runaway stars span the entire range of masses in the sample and have masses as low as 15~$\msun$. We find space velocity is independent of spectral type. Therefore, the underlying cause as to why low-velocity stars have low mass-loss rates cannot simply be attributed to the kinematic properties of these stars. $V_{2D}$ being independent of stellar mass strongly suggests that we must consider the effects of ISM bulk flows for the slowest stars in our sample, per the suggestion of \citet{vandenEijndelandKaper2026}.


\begin{figure}[htb!]
        \plotone{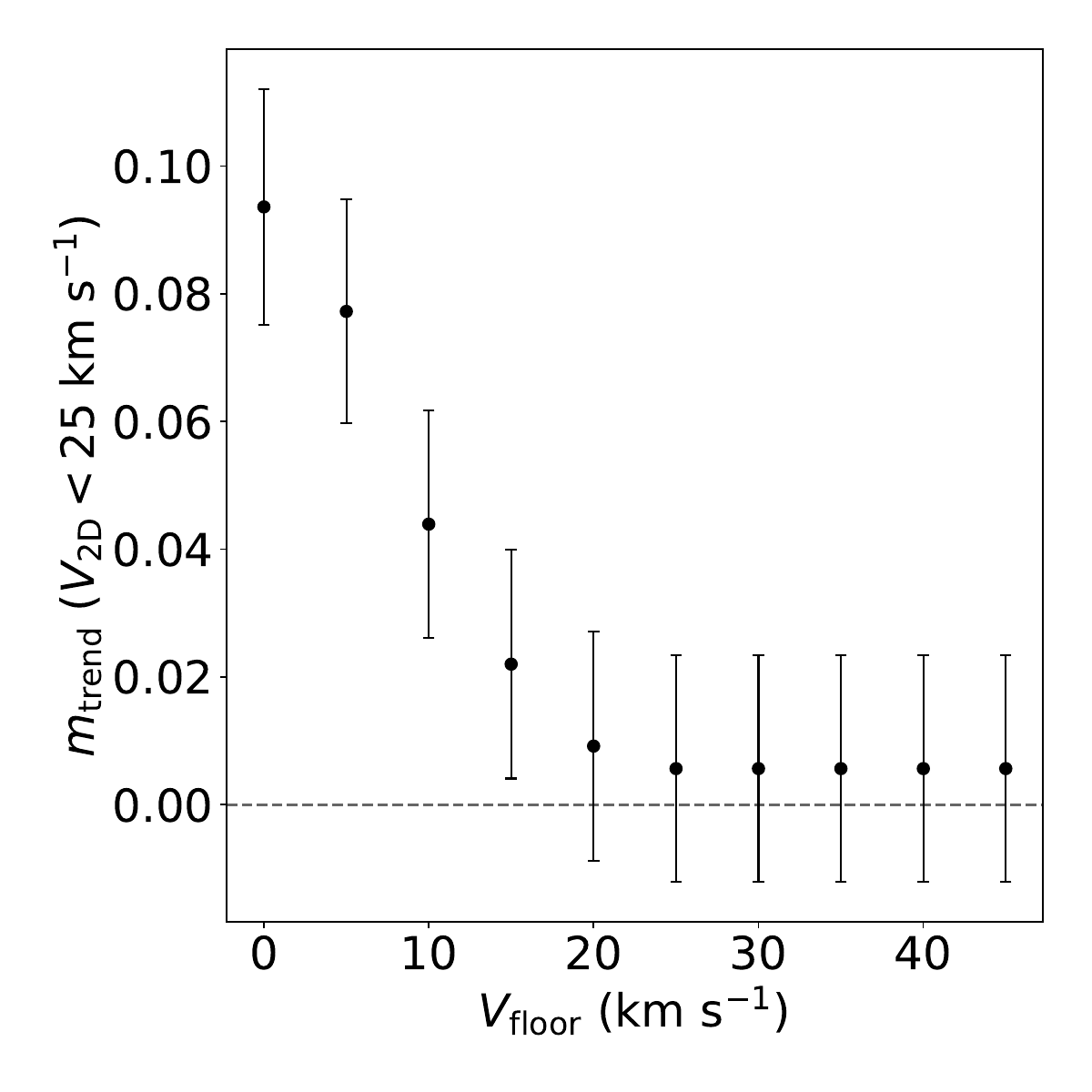}
        \caption{The slope between $\dot{M}$ and $V_{2D}$ for low-velocity stars (as in Figure~\ref{fig:v_mdot_correlation}) for different velocity floors. Error bars show the uncertainty in the computed slopes and the dashed-horizontal line shows zero slope.}
        \label{fig:v_mdot_correction}
\end{figure}

We investigate the correlation in $\mdot$ and $V_{2D}$ by adopting several lower-velocity limits. For each lower-velocity limit, each star's velocity was corrected to a minimum value and $\mdot$ was recomputed for each case. We then computed the slope of the correlation between $\mdot$ and $V_{2D}$ to investigate how the slope of the trend line changes with varying $V_{2D}$ floors. Figure~\ref{fig:v_mdot_correction} plots the slope and uncertainty of the best-fit line between $\mdot$ and $V_{2D}$ (for low-velocity stars) versus the adopted minimum-velocity value. We find the correlation between $\mdot$ and $V_{2D}$ disappears when a minimum velocity of 20~$\kms$ is adopted. This is in good agreement with the findings of \citet{vandenEijndelandKaper2026}, who found bulk flows of 15~$\kms$ must be responsible for producing bow shocks around low-velocity stars. Based on the findings of \citet{vandenEijndelandKaper2026} and the evidence of Figure~\ref{fig:v_mdot_correction}, we are physically motivated to adopt a velocity floor of $V_a=20~\kms$ for stars with space velocities less than this value. This lower limit is also consistent with numerical simulations of bow shocks \citep{ComeronandKaper1998, Meyer2016, Acreman2016}, which find that relative velocities of 20~$\kms$ is representative of the slowest bow shock-forming stars.

\begin{figure*}[htb!]
        \plotone{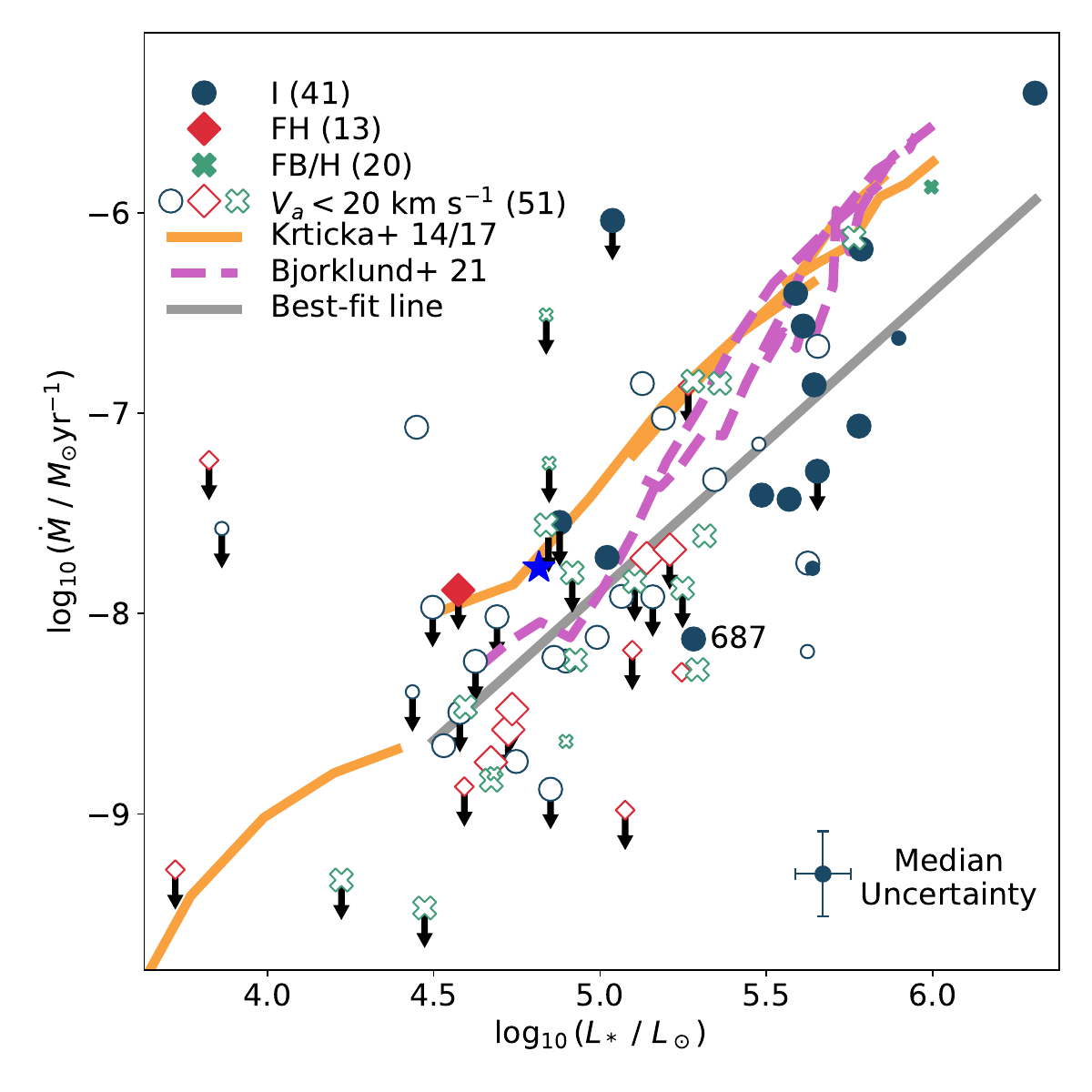}
        \caption{Mass-loss rates corrected to a minimum velocity of $V_a=20~\kms$ versus stellar luminosity. Points are colored in the same way as in Figure~\ref{fig:mass_loss_full}. The grey track signifies best-fit trend line for measurement of $\mdot$ for stars above $\log\lstar=4.5$. Hollow points have shifted upwards compared to Figure~\ref{fig:mass_loss_full}. With this correction, our measurements show good agreement with the \citet{Bjorklund2021} models.}
        \label{fig:mass_loss_floor}
\end{figure*}




Figure~\ref{fig:mass_loss_floor} plots the logarithm of the corrected mass-loss rate $\dot{M}_{20}$---adopting a minimum relative velocity of 20~$\kms$---versus the logarithm of stellar luminosity, in solar luminosity units, as in Figure~\ref{fig:mass_loss_full}. Hollow points again denote stars with low measured space velocities, but now these are assumed to have $V_a=20~\kms$. These stars have shifted upward substantially relative to their positions in Figure~\ref{fig:mass_loss_full} and have joined the trend defined by the high-confidence, high-velocity stars. As before, the points show a strong correlation between $\mdot$ and $\lstar$. The grey best-fit line shows excellent agreement with the predictions and is coincident to the \citet{Bjorklund2021} models, at least for the lower-luminosity stars. This is remarkable because there are no free parameters in the mass-loss rate calculation. Even at the high-luminosity end, the trend line is only offset downward from the predictions by about 0.3~dex. The dispersion of the data from the mean relation is 0.58~dex, which is 0.27 dex smaller than the RMS in Figure~\ref{fig:mass_loss_full}. Low-velocity stars with assumed $V_a=20~\kms$ show a similar degree of scatter from the mean relation as the high-velocity stars (0.60~dex versus 0.56~dex). Low-velocity stars with no detectable 70~$\um$ signal (\textit{i.e.}, hollow points with downward-facing arrows) now scatter both above and below the models, as opposed to beneath the models as in Figure~\ref{fig:mass_loss_full}. This would seem to indicate that the systematic downward shift of low-velocity stars in Figure~\ref{fig:mass_loss_full} has been properly corrected here, but the correction may be insufficient for some individual cases.



BS687$_\star$ now appears as an outlier below the theoretical predictions and the general trend in the other data near the center of the plot. The star has a peculiar 2D space velocity of 20~$\kms$, which is just high enough to not be affected by our $V_a$ correction. Since the measurement of $\mdot$ is anomalously low for its luminosity, and its $V_a$ is just above the threshold for the correction, it is likely that this star also experiences an ISM bulk flow which causes the total relative velocity beeween the star and its surroundings $V_a$ to exceed the correction value of 20~$\kms$. This is consistent with outflows from \ion{H}{2} regions \citep{Bodenheimer1979}. This highlights the potential problems of applying a constant minimum velocity correction, as ISM bulk flows can exceed the correction we apply. Ideally, we would apply a correction for ISM bulk flows on a case-by-case basis, but we must apply a minimum average correction owing to the infeasibility of measuring ISM bulk flows for every star.



\subsection{The bi-stability jump}

\begin{figure*}[htb!]
        \plotone{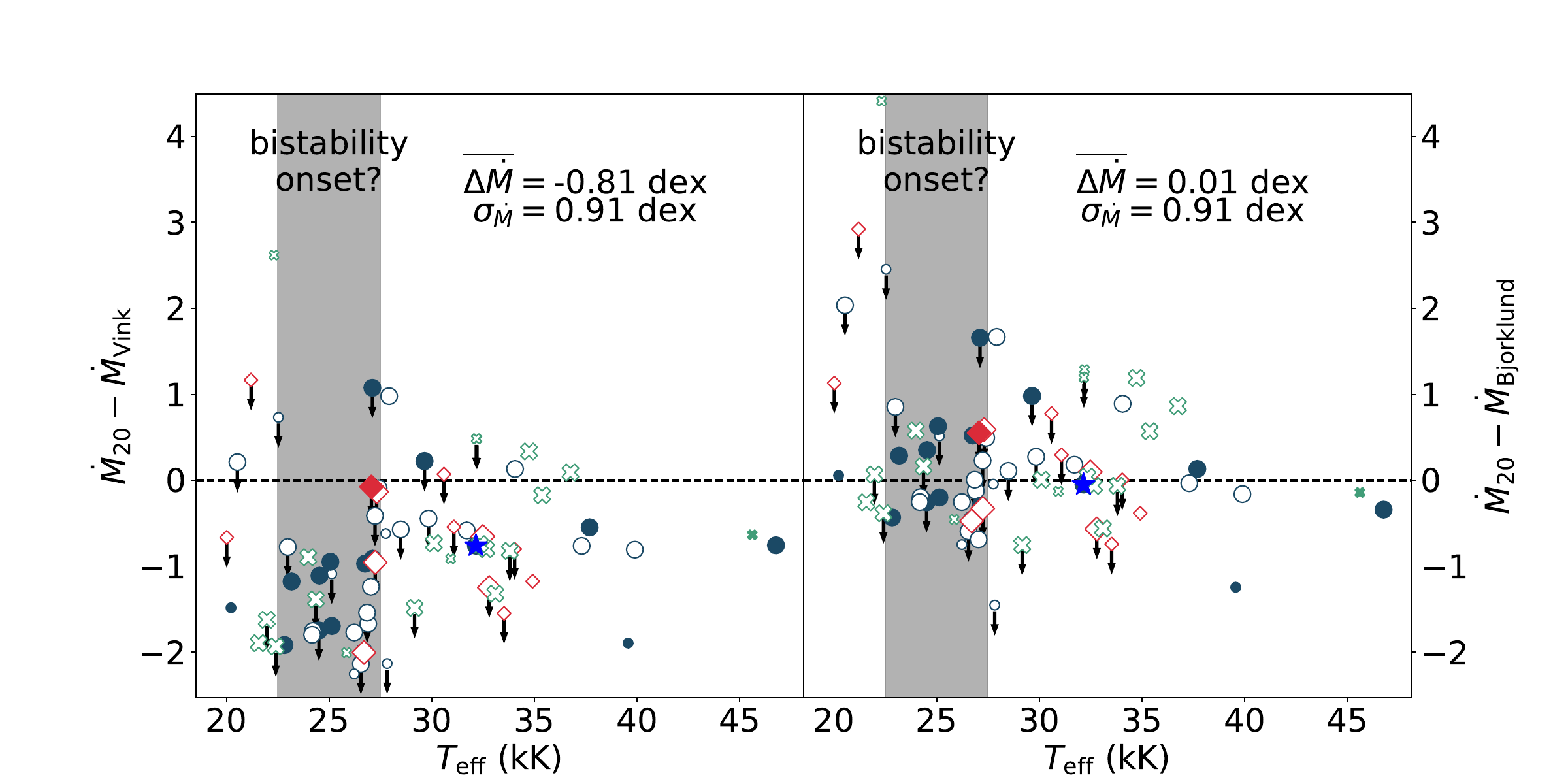}
        \caption{The difference between corrected mass-loss rates $\mdot_{20}$ and the V01 prediction (left panel) and the difference between corrected mass-loss rates $\mdot_{20}$ and the B23 prediction (right panel) versus stellar effective temperature. Our corrected measurements of $\mdot$ show better agreement with the B23 models over the V01 prediction, suggesting that $\mdot$ measured from the bow shock technique does not show signs of bi-stability enhancements.}
        \label{fig:mass_loss_temp_diff}
\end{figure*}

We search for signatures of enhanced $\mdot$ due to bi-stability behavior near 25~kK by plotting in Figure~\ref{fig:mass_loss_temp_diff} the difference between corrected mass-loss rates ($\mdot_{20}$) and predictions from the analytical formulae of V01 (which includes bi-stability enhancements, left panel) and of B23 (which does not include bi-stability enhancements, right panel) versus stellar effective temperature in kK. The gray shaded region in both panels shows where the predicted bi-stability behavior should appear in the 22.5--27.5~kK range. The horizontal dashed line marks a difference of zero. 


We find the B23 formula matches our corrected measurements of $\mdot$ better than the V01 prescription. The points in the left panel are shifted---on average---0.81~dex below the dashed line, meaning our measurements are systematically lower than what Equations~24 and 25 from V01 predict. In comparison, points in the right panel are offset 0.01~dex above the B23 prediction, which is small in comparison to the median uncertainty in $\mdot_{20}$. Many observational studies have suggested a reduction in the V01 relation by a factor of 2--3 \citep[][]{Najarro2011, Sundqvist2011, Bouret2012}, and our results are in agreement with this finding. The points in both panels have a dispersion of 0.91~dex about the mean, which is 4 times greater than the median statistical uncertainty in our corrected $\mdot$. Points in the left-hand panel also show a systematic deviation from the V01 relation, with cooler stars lying farther below the prediction than hotter stars, while no such trend is evident in the right-hand panel. Our data show no effect of bi-stability enhancements, consistent with the findings of recent studies \citep{deBurgos2024, Alkousa2025}. The four outliers at the upper left in the right-hand panel are upper limits and therefore, do not constitute evidence of enhanced mass loss from bi-stability behavior.

Interestingly, K19 claimed evidence of bi-stability in the previous application of the bow shock technique in their study of 70 bow shock stars, with stars on the cool side of the bi-stability gap having mass-loss rates 1--4 dex higher than the V01 prescription. This result was surprising, as V01 already includes bi-stability effects, and these findings suggested that bi-stability enhancements may have been more significant than predicted since the bow shock technique was able to probe less-luminous stars which would otherwise be inaccessable by other techniques. Focusing on the coolest stars in our sample, our findings run counter to those of K19.

To investigate this discrepancy, we turn our attention to stars cooler than 25~kK that are in common in both this work and the sample of K19, which yields 4 stars---BS$007_\star$, BS$357_\star$, BS$361_\star$, and BS$362_\star$. We use nearly identical values for all parameters in Equation~\ref{eq:two}, except for $I_\nu$. For all stars, our values of $I_\nu$ are at least a factor of four lower, and as much as a factor of six lower. This can largely be attributed to the erroneous pixel size used in K19. We attribute the rest of this difference to our more rigorous methodology by which we fit a profile to the 70~$\um$ data and subtract an estimate of the background. Both differences inflated $\mdot$ for these cool stars in K19, and likely the other stars cooler than 25~kK in that work, giving a false impression of bi-stability enhancements. In some cases, we determine objects are not detected at 70~$\um$, in disagreement with K19. Our corrected measurements of $\mdot$ agree well with the analytical formulae of B23, and we take this as evidence of the lack of bi-stability enhancements.



\subsection{Enhancement in $\mdot$ for rapid rotators?}

\begin{figure}[htb!]
        \plotone{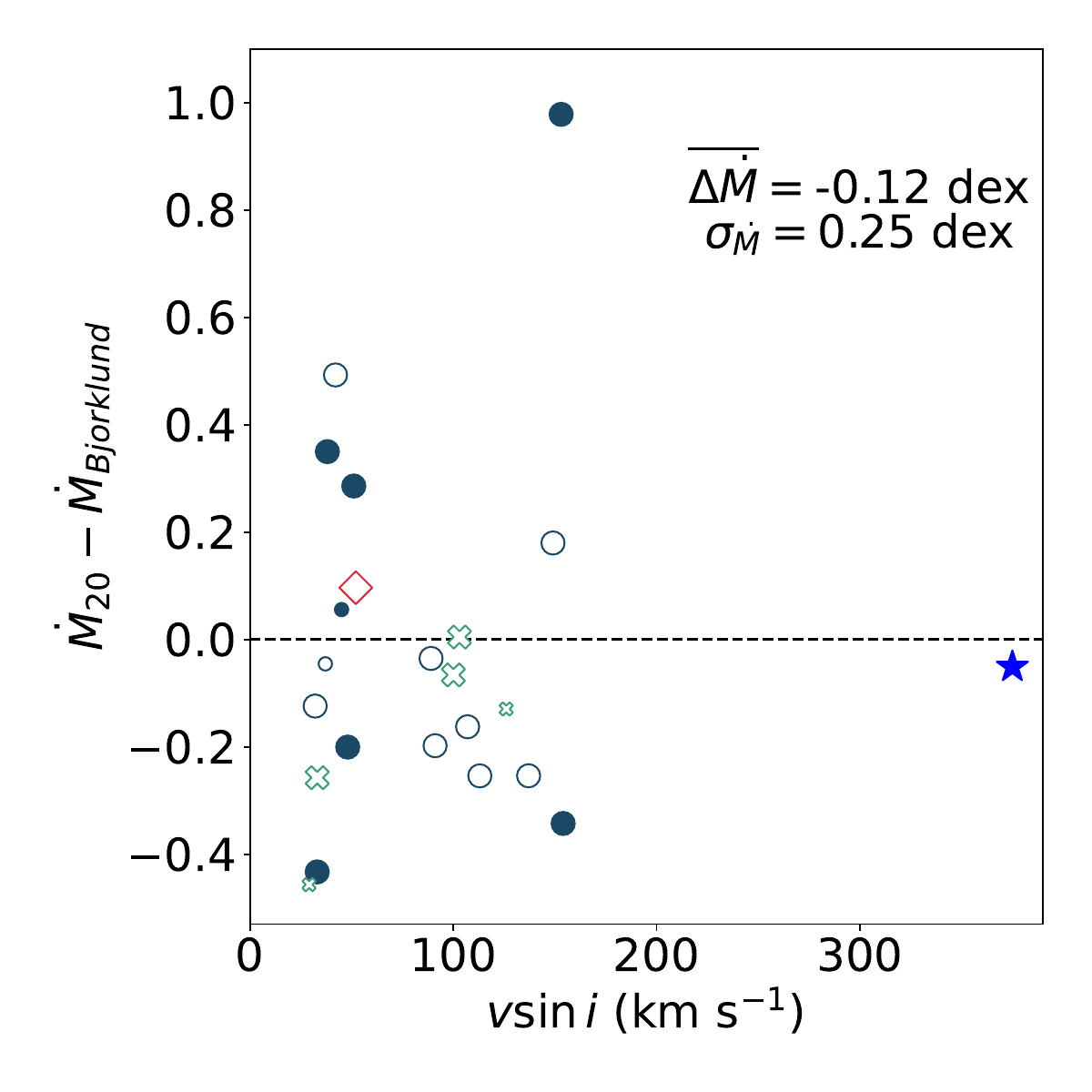}
        \caption{The difference between the corrected mass-loss rates $\mdot_{20}$ and the B23 prediction versus projected rotational velocity. There is no correlation, suggesting that any effects from rotation on mass loss are small compared to the uncertainties.}
        \label{fig:mass_loss_diff_vsini}
\end{figure}

Figure~\ref{fig:mass_loss_diff_vsini} plots the difference between the corrected mass-loss rates of bow shock stars, $\mdot_{20}$, and the B23 prediction (which was computed with a rotation rate 40\% the critical rotational velocity) versus the projected rotational velocity $v\sin i$. Points are colored the same as in Figures~\ref{fig:mass_loss_full} and \ref{fig:mass_loss_floor}. We exclude bow shock stars that we classify as double-lined spectroscopic binaries in P25 as well as bow shock upper-limit cases, which leaves 23~stars. The points show a uniform dispersion about the horizontal dashed line, with a median difference from the B23 prescription of -0.12~dex---which is smaller than the median uncertainty in $\mdot_{20}$---and a standard deviation of 0.25~dex. $\zeta$~Oph, the fastest rotator in our sample at $v\sin i=360~\kms$, is an illustrative case and shows excellent agreement with the B23 prediction. $\zeta$~Oph---which is near its critical rotational velocity (see Figure~13 of P25)---should show the most dramatic increase in mass-loss rate if rotation is a significant factor in determining mass loss. From the limited number of rapidly rotating stars available, we conclude that any effects of rotational velocity on $\mdot$ are minimal compared to the random uncertainties in measuring $\mdot$ using the bow shock technique.

\section{Discussion \& Conclusions}
\label{sec:discussion}






We have shown that the bow shock method is capable of measuring mass-loss rates for stars with luminosities $4.5\lesssim\log\left(\lstar/L_\odot\right)\lesssim6.3$. This includes stars less luminous than the weak-winded star $\zeta$~Ophiuchi. Our sample contains both dwarf and evolved stars. At $\log\left(\lstar /L_\odot\right)\lesssim4.5$, there are very few objects that can be studied, but even in this regime the bow shock technique is promising, since late-O and early B stars are largely inaccessible by other techniques of measuring mass loss. Many of these low-luminosity stars have $\mdot$ upper limits, and a limiting factor for this method is the detectability of the bow shock above the 70~$\um$ background. Overlapping foreground and background emission at 70~$\um$ dilutes the signal from the bow shock nebula and inflates mass-loss rates above predictions in some cases. However, taken as a whole, these upper-limit cases are close to the overall trend. This may signify that some upper limits are nearly detections, but the signals are still not strong enough to distinguish over the background.

Our measurements of $\mdot$ lie below the analytical mass-loss prescriptions of V01 by 0.83~dex, but they show excellent agreement with the predictions of B23---with a median offset of +0.01~dex. We do not find evidence of bi-stability effects, which runs counter to the previous bow shock work of K19, but is in line with other recent observational studies \citep{deBurgos2024, Alkousa2025}. Overall, we find a reduction in mass-loss rates from the previous bow shock studies of K18/K19 due to the use of the correct pixel size and more-rigorous background subtraction when measuring $I_\nu$.




Previous works have shown that most stars that form bow shock nebulae are not runaways and have suggested that interstellar flows play a role in forming the bow shock. Mass-loss rates computed by taking the peculiar-space velocities as the star-ISM relative velocity show an unexpected and unphysical systematic correlation between $\mdot$ and $V_a$ for slow-moving stars. This motivated the introduction of a floor on $V_a$ of 20~$\kms$---representative of the true star-ISM velocity---which is corroborated by the work of \citet{vandenEijndelandKaper2026}. This is a large but physically motivated correction (e.g., Figure~\ref{fig:mass_velocity_correlation}). When this correction is applied, we find it produces a relation with a smaller RMS (0.58~dex versus 0.85~dex) about the trend and a relation that is in better agreement with models. This correction to $V_a$ is necessary for the use of this method of obtaining $\mdot$. The uncertainty in $V_a$ is the dominant source of error in $\mdot$ but nevertheless, this method is successful in producing a tight relationship between mass-loss rate $\mdot$ and luminosity $\lstar$, with the same slope and offset as models. With the recognition of the need for the lower limit on $V_a$, the bow shock method has become a successful technique for measuring $\mdot$ with striking agreement to modern models. The dispersion in this relation is small compared to radio-recombination, H$\alpha$, and UV-based studies. For example, the dispersions in $\mdot$ as found in \citet{HowarthandPrinja1989} were about a full dex. Recent compilations yield measurements of $\mdot$ with dispersions of about 0.5~dex \citep{deBurgos2024}, similar to this work.

The results for objects within \ion{H}{2} regions, facing \ion{H}{2} regions, and facing bright-rimmed clouds, are consistent with measurements from isolated nebulae. We conclude that this technique is robust for bow shocks in any of these environments. This makes the bow shock technique applicable to a wide range of scenarios in which stellar bow shock nebulae can form and does not require knowledge about the immediate environment. This work establishes the bow shock technique as a promising and robust method with no adjustable parameters---with the exception of the physically motivated lower limit on $V_a$. The primary limitation on broad use of this method is the limited number of 70~$\um$ detections.

\begin{acknowledgments}

This work has been supported by the National Science Foundation through grants AST-2108348, AST-1852289, and AST-2108349. We thank Stan Owocki, whose contribution to the field of stellar outflows is a great inspiration to us \citep{Stanfest}. We also thank Daniel Maschmann for his coding expertise. We thank Dr. Matthew Povich for helpful comments on early versions of this paper.

This work is based, in part, on observations made with the Spitzer Space Telescope, which was operated by the Jet Propulsion Laboratory, California Institute of Technology under a contract with NASA. Support for this work was provided by NASA through an award issued by JPL/Caltech.
\end{acknowledgments}






%
\facilities{Gaia, Herschel (PACS), Spitzer (IRAC, MIPS), WISE}


\software{astropy \citep{2013A&A...558A..33A,2018AJ....156..123A,2022ApJ...935..167A}
          }



\bibliography{bibliography}{}
\bibliographystyle{aasjournalv7}



\end{document}